\documentclass{aa}      
\usepackage{upgreek} 
\usepackage{color}

\usepackage{chemformula}
\usepackage{mhchem}
\usepackage{amsmath}

\usepackage{hhline}
\usepackage{longtable} 
\usepackage{graphicx}
\usepackage{tabularx}   

\usepackage[varg]{txfonts}
\usepackage{upgreek} 
\usepackage{multirow} 
\usepackage[FIGTOPCAP, center, nooneline]{subfigure}
\usepackage[]{natbib}

\usepackage[varg]{txfonts}
\usepackage{microtype}
\usepackage{hyperref}
\usepackage{booktabs}
\usepackage{xcolor}
\usepackage{hyperref}
\usepackage{graphicx}
\usepackage{multirow}
\usepackage{float}

\hypersetup{%
  colorlinks=true,
  linkcolor=blue,
  citecolor=blue
}

\usepackage{etoolbox}

\let\oldcitet=\citet

\renewcommand{\citet}[1]{\textcolor[rgb]{0,0,1}{\oldcitet{#1}}}

\newcommand{\FeII}{Fe\,{\sc ii}}

\newcommand{\blue}[1]{\textcolor{blue}{#1}}

\begin{document} 

\title{H$_{3}^{+}$ in irradiated protoplanetary  disks:\\Linking far-ultraviolet radiation and  water vapor}

\titlerunning{FUV-driven formation of H$_{3}^{+}$ in irradiated disks} 
\authorrunning{Goicoechea et al.}

 \author{Javier R.\,Goicoechea\inst{1}
          \and
    Octavio\,Roncero\inst{1}
    \and
    Evelyne Roueff\inst{2}
    \and
    John H. Black\inst{3}
    \and
   Ilane Schroetter\inst{4}    
    \and
    Olivier Bern\'e\inst{4} 
     }
  
\institute{Instituto de F\'{\i}sica Fundamental
     (CSIC). Calle Serrano 121-123, 28006, Madrid, Spain. \email{javier.r.goicoechea@csic.es}
\and
LUX, Observatoire de Paris, Universit\'e PSL, Sorbonne Universit\'e, CNRS, 92190, Meudon, France.
\and
Chalmers University of Technology, Onsala Space Observatory, 
Onsala, Sweden.
\and
Institut de Recherche en Astrophysique et Plan\'etologie, Universit\'e de Toulouse, CNRS, CNES, Toulouse, France.}

   \date{Received 6 June 2025 / Accepted 20 September 2025}

\abstract{The likely JWST detection of vibrationally excited H$_3^+$ emission in Orion's irradiated disk system d203-506 raises the important question of whether cosmic-ray ionization is enhanced in disks within clustered star-forming regions, or whether alternative mechanisms contribute to H$_3^+$ formation and excitation.
We present a detailed  model of the photodissociation region (PDR) component of a protoplanetary disk --comprising the outer disk surface and the photoevaporative wind -- exposed to strong external far-ultraviolet (FUV) radiation.
We investigate  key gas-phase reactions involving excited H$_2$ that lead to the formation of H$_3^+$ in the PDR, including  detailed state-to-state dynamical calculations of  reactions \mbox{H$_2$($v$\,$\geq$\,0)\,+\,HOC$^+$\,$\rightarrow$\,H$_3^+$\,+\,CO} and 
\mbox{H$_2$($v$\,$\geq$\,0)\,+\,H$^+$\,$\rightarrow$\,H$_2^+$\,+\,H}.
We also consider the effects of photoionization of vibrationally excited H$_2$($v \geq 4$), a process not previously included in PDR or disk  models.
We find that these \mbox{FUV-driven} reactions dominate the formation of H$_3^+$ in the PDR of strongly irradiated disks,
largely independently of cosmic-ray ionization. 
The predicted H$_3^+$ abundance in the disk PDR peaks at $x(\mathrm{H}_3^+) \gtrsim 10^{-8}$, coinciding with regions of enhanced HOC$^+$ and water vapor abundances, 
and is linked to the strength of the external FUV field ($G_0$). 
The predicted H$_3^+$ column density ($\lesssim 10^{13}\,\mathrm{cm}^{-2}$) 
agrees with the presence of H$_3^+$ in the PDR of d203-506.
We also find that formation pumping, resulting from exoergic reactions between excited H$_2$ and HOC$^+$,  drives the vibrational excitation of
H$_3^+$ in these regions. We expect this photochemistry to be highly active in disks where $G_0 > 10^3$.
The H$_3^+$ formation pathways studied here may also be relevant in the inner disk region  (near the host star), in exoplanetary ionospheres, and in the early Universe.
}

\maketitle
%

\section{Introduction}\label{sec:introduction}

Most protoplanetary systems form in stellar clusters \mbox{\citep{Lada03}} and
are therefore  eventually exposed to external 
ultraviolet (UV) radiation from nearby massive stars \mbox{\citep[e.g.,][]{Winter22}}. 
UV irradiation can influence the disk’s chemical composition \citep{Walsh13,Boyden20,Boyden23,Berne23,Berne24}, impact gas line excitation \citep{Zannese24,Goico24}, and alter the potential for planet formation.
In particular, external UV radiation leads to disk mass loss driven by photoevaporative  winds \citep{Johnstone98,Stoerzer99}, which truncate the outer disk, reduce the mass reservoir for planet formation, and  impact the properties of gas giant planets forming within those disks
\citep[e.g.,][]{Winter22b,Qiao23,Huang24}. 
However, the extent to which environmental UV radiation affects chemical processing and gas ionization remains controversial \citep[e.g.,][]{Ramirez23,Diaz24,Ilane25a}. This is a  relevant question in the context of Solar System formation, as evidence suggests that the early solar nebula disk formed in a stellar cluster, not far from a massive star 
\citep[][]{Adams10,Bergin23,Desch24}.

Massive OB-type stars release intense UV radiation, reaching the disk population that gradually emerges from the natal cloud 	\citep[e.g.,][]{Terwisga23}. Extreme-UV 
(\mbox{EUV; 13.6\,$<$\,$E$\,$<$\,100\,eV}) photons ionize the photoevaporative wind, creating the characteristic ionization front teardrop morphology around  
\mbox{proplyds} \citep[e.g.,][]{ODell93,Bally00,Mauco23,Aru24}. \mbox{External} \mbox{far-UV} (FUV; \mbox{6 $<$\,$E$\,$<$\,13.6\,eV}) 
 photons  penetrate deeper into the disk, heating the gas and dust and altering its chemical composition by ionizing atoms with low ionization potential (IP), triggering the formation of molecular ions and radicals 
 \citep[e.g.,][]{Berne23,Berne24,Zannese24,Zannese25}, and desorbing ice grain mantles
 \citep[e.g.,][]{Walsh13,Goico24}. This \mbox{FUV-irradiated}  disk zone, a \mbox{``photodissociation region''} \citep[PDR, e.g.,][]{Hollenbach99}
   exhibits high ionization fractions
 ($x_e$\,$=$\,$n(\rm e^-)$/$n_{\rm H}$), a fundamental parameter in  disk chemistry and dynamics, with
 $x_e$ of a few 10$^{-4}$ at the PDR edge if $x_e$\,$\approx$\,$x_{\rm C^+}$.  
 Cosmic rays have a greater penetration depth. They are responsible for the much lower, but somewhat uncertain, ionization fraction  in the \mbox{FUV-shielded}  disk midplane, with $x_e$\,$\propto$$\sqrt{\zeta/n_{\rm H}}$ down to \,$\approx$10$^{-11}$, where $\zeta$ is the H$_2$ ionization rate.  Here, we assume that $\zeta$ is dominated by cosmic-ray ionizations, though it may also have local contributions from stellar X-rays and the decay of short-lived radionuclides \citep[e.g.,][]{Cleeves13,Cleeves15}.

Identifying tracers of FUV, $x_e$, and $\zeta$ in various regions of a protoplanetary disk is critically important.
Since its  detection in the interstellar medium (ISM)  of our galaxy \citep{Geballe96} and beyond \citep{Geballe06}, the   IR ``absorption'' spectrum of H$_{3}^{+}$ is considered one of the cleanest probes \mbox{of $\zeta$} 
\citep[e.g.,][]{McCall02,Goto08,Indriolo12}.
In ISM clouds, cosmic-ray ionization of H$_2$ produces H$_2^+$, which rapidly reacts with 
molecular hydrogen to form H$_3^+$, 
\begin{equation}
{\rm H_2^+}\,+\,{\rm H_2}\,\rightarrow\,{\rm H_3^+}\,+\,{\rm H}.
\label{reaction_h3p}
\end{equation}
This is a  highly exothermic reaction, $\sim$1.7\,eV ($\sim$20{,}000\,K), nearly insensitive to
the initial vibrational state of \ch{H2+} for excitation energies below $\sim$1\,eV
\citep{delMazo-Sevillano-etal:24a}.
 Depending on \( x_e \), the destruction of H$_3^+$ can be dominated either by dissociative recombination with electrons or by exothermic proton transfer reactions with abundant species  such as CO, N$_2$, and O 
\citep[e.g.,][]{Herbst73,Watson73,Black77}. The likely detection\footnote{Although H$_3^+$ possesses equilateral triangle symmetry, and therefore lacks a permanent dipole, its vibrational modes can break this symmetry, leading to strong dipole-induced transitions \citep[e.g.,][]{Tenysson94,Miller20}. The only fully allowed transition is the infrared-active  asymmetric stretch-bend mode $\nu_2$, with a band origin at  \mbox{2,521.6 cm$^{-1}$} ($\sim$4\,$\mu$m), first observed in the laboratory by \cite{Oka80}.} of infrared (IR) H$_{3}^{+}$ 
\mbox{``emission''} \citep[see][]{Ilane25b} in the irradiated  disk \mbox{d203-506}, 
in the line of sight toward the  Orion Bar,  raises the question of whether $\zeta$ is \mbox{enhanced} in  disks within clustered star-forming regions
or whether alternative mechanisms contribute to H$_3^+$ formation.

In contrast to interstellar  absorption  measurements,  H$_{3}^{+}$ emission has been detected in Jupiter's ionosphere through observations of its 
\mbox{$\nu_2=2\rightarrow 0$} \mbox{overtone} at $\sim$2\,$\mu$m  \citep{Drossart89}
and its fundamental \mbox{$\nu_2=1\rightarrow 0$} band at $\sim$3.4--4.4\,$\mu$m
\citep{Oka90}.  
  The detection of H$_{3}^{+}$  emission in Jupiter, Saturn \citep{Geballe93}, Uranus \citep{Trafton93}, and Neptune \citep{Melin25} probes solar wind-driven \mbox{ionospheric} and auroral activity.
The main source of H$_2$ ionization in  planetary ionospheres is high-energy electron impacts, \mbox{${\rm H_2}\,+\,{\rm e^{-\,*}}\,\rightarrow\,{\rm H_2^+}\,+\,{\rm 2e^-}$},
 with some contribution from ionization by solar EUV radiation \cite[e.g.,][]{Miller00,Miller20}.

H$_{3}^{+}$ line emission outside the Solar System was first (tentatively) reported  in the ejecta of Supernova 1987A \citep[][]{Tenysson94,Lepp91}  and in the  disk around the Herbig Ae/Be star HD 141569 \citep{Brittain02}. The later detection was disputed by \cite{Goto05}, who did not find any  H$_{3}^{+}$ lines in follow-up observations. Until now, the most robust detection of interstellar H$_{3}^{+}$ \mbox{$\nu_2=1\rightarrow 0$}
emission has come from JWST observations of  ultraluminous IR galaxies 
\citep[ULIRGs,][]{Pereira24}. The detection of ro-vibrational emission raises questions about 
 H$_3^+$  formation and excitation mechanisms   \citep[collisional, radiative, or formation pumping; see e.g.,][]{Anicich84,LeBourlot24}.

A key feature of dense \mbox{FUV-irradiated} gas is the presence of significant quantities of thermal and \mbox{FUV-pumped} rotationally and vibrationally excited\footnote{\citet{Berne24} reported the detection of
IR H$_2$ lines up to $v$\,=\,4 originating from the PDR component of \mbox{d203-506}.
They also detected highly excited rotational lines of  \mbox{H$_2$($v=0$) up to
$J=17$.}
 Higher spectral resolution observations of interstellar PDRs such as the Orion Bar show H$_2$ lines in vibrationally excited levels up to $v$\,=\,12  \citep{Kaplan21}. } \mbox{H$_2$}, hereafter H$_2^*$, which is typically much more chemically reactive than \mbox{H$_2$($v$\,=\,0,\,$J$\,=\,0)}, the dominant form of H$_2$ in cold interstellar gas \citep[][]{Stecher72,Freeman82,Tielens85,Sternberg95,Agundez10}. 
In PDRs, H$_2^*$ initiates  a  ``hot'' chemistry, whereby endoergic reactions (endothermic or with activation energy barriers) become active and proceed rapidly.
This \mbox{``state-dependent''} chemistry in disks remains poorly characterized \citep[e.g.,][]{Walsh13,Ruaud21}, as few reaction rates are known and advanced computational methods are required to determine these rate constants \citep[e.g.,][]{Zanchet13a,Zanchet13b,Zanchet19,Veselinova21,Goico22b}. 
In this study we investigate the chemical pathways through which strong FUV radiation fields trigger the formation of abundant H$_3^+$ via a few specific reactions  involving H$_2^*$.
We theoretically investigate two fundamental reactions:
\mbox{${\rm H_2}(v'')\,+\,{\rm H^+}\,\leftrightarrow\,{\rm H_2^+}(v')\,+\,{\rm H}$},
which is very endoergic in the left to right direction when \mbox{$v''$\,$<$\,4} (by $\sim$21,200\,K when \mbox{$v''$\,=\,0}),
 and reaction 
\mbox{${\rm H_2}(v'')\,+\,{\rm  HOC^+}\,\rightarrow\,{\rm H_3^+}(v')\,+\,{\rm CO}$}, 
which is endoergic by $\sim$1{,}500\,K  when \mbox{$v''$\,=\,0} 
and \mbox{$J''$\,=\,0} (see \mbox{Appendices}).
Using  detailed dynamical computations,  we determine their vibrational-state dependent  rate constants and incorporate them into our  model
simulating the PDR component of \mbox{d203-506}. In addition, we investigate the role of \mbox{FUV-photoionization} of  H$_2^*$($v \geq 4$)  \citep{Ford75}, another state-dependent process that leads to the formation of H$_2^+$, and consequently H$_3^+$.

In \mbox{Sect.~\ref{sec:intro_d203-506}}, we summarize the most relevant properties of \mbox{d203-506}. In \mbox{Sect.~\ref{sec:PDR_reference}}, we present our reference model of the disk's PDR component. In \mbox{Sect.~\ref{sec:chemistry}}, we dissect this component into three zones and study the  dominant chemical formation pathways for H$_3^+$ in these zones. In \mbox{Sect.~\ref{sec:discussion}}, we present and discuss our photochemical and excitation models.

\section{The  irradiated  disk d203-506 and its H$_{3}^{+}$ emission} 
\label{sec:intro_d203-506}

The  disk 
\mbox{d203-506} 
was first detected in silhouette against the optical background of the Orion Nebula 
by  \mbox{\textit{Hubble}}	\citep{Bally00} and later in HCO$^+$
by ALMA \mbox{\citep{Champion17}}.
More recent observations from the Keck Observatory, Very Large Telescope (VLT), and 
\textit{James Webb} Space Telescope (JWST) have revealed the extended nature of its PDR component -- the irradiated outer disk surface and  photoevaporative wind -- created by FUV radiation from \mbox{$\theta^1$ Ori C}, the most massive O-type star in the Trapezium 
\citep{Berne23,Berne24,Haworth23,Habart24,Goico24,Zannese24,Zannese25}.

While the exact  position (in the line of sight) of d203-506 is uncertain 
\mbox{\citep[e.g.,][]{Haworth23}}, the Orion Nebula Cluster lies at a distance of $\sim$400\,pc \citep[e.g.,][]{Kounkel18}. \citet{Berne24} determined 
 the mass of the host star 
($\sim$0.3\,M$_{\odot}$), and the disk's mass (\mbox{$\sim$\,10\,M$_{\rm Jup}$\,$\simeq$\,10$^{-2}$\,M$_{\odot}$}) and  radius \mbox{($\sim$100\,au\,$\simeq$\,0.25$''$)}.
The estimated flux of the external FUV field, derived from near-IR fluorescent
OI and CI emission lines, is $G_0$\,$\simeq$\,2$\times$10$^4$ 
\citep{Berne24,Goico24}, where $G_0$ is the mean interstellar FUV field in Habing units 
\citep[$G_0$\,=\,1 is equal to \mbox{1.6$\times$10$^{-3}$\,erg\,cm$^{-2}$\,s$^{-1}$};][]{Habing68}.
In addition, \mbox{d203-506} exhibits shock-excited gas from a small jet (HH\,520), observed in [\FeII]  \citep{Haworth23,Berne24}.  
Most molecules exhibiting extended emission in the disk's PDR show their  peak near a bright ``spot'' facing the Trapezium 
\citep{Berne24}, which likely corresponds to a column density enhancement, perhaps  the interaction region between the PDR and the jet.  

This disk  is particularly interesting for constraining the effects of  FUV radiation, as it does not exhibit ionization fronts produced by EUV radiation \citep{Haworth23}. 
In addition, the nearly edge-on orientation of \mbox{d203-506}, with an 
estimated inclination of $i>75^{\circ}$ \citep[based on the non-detection of the internal star;][]{Berne24}, makes it an excellent target for studying the vertical effects of FUV radiation in the upper disk and neutral wind.
This geometry allows for the PDR component to be isolated more easily, since the very high dust optical depth at short wavelengths renders IR observations insensitive to the cold disk midplane and inner disk.

\begin{figure}[h!]
\centering   
\includegraphics[scale=0.36,angle=0]{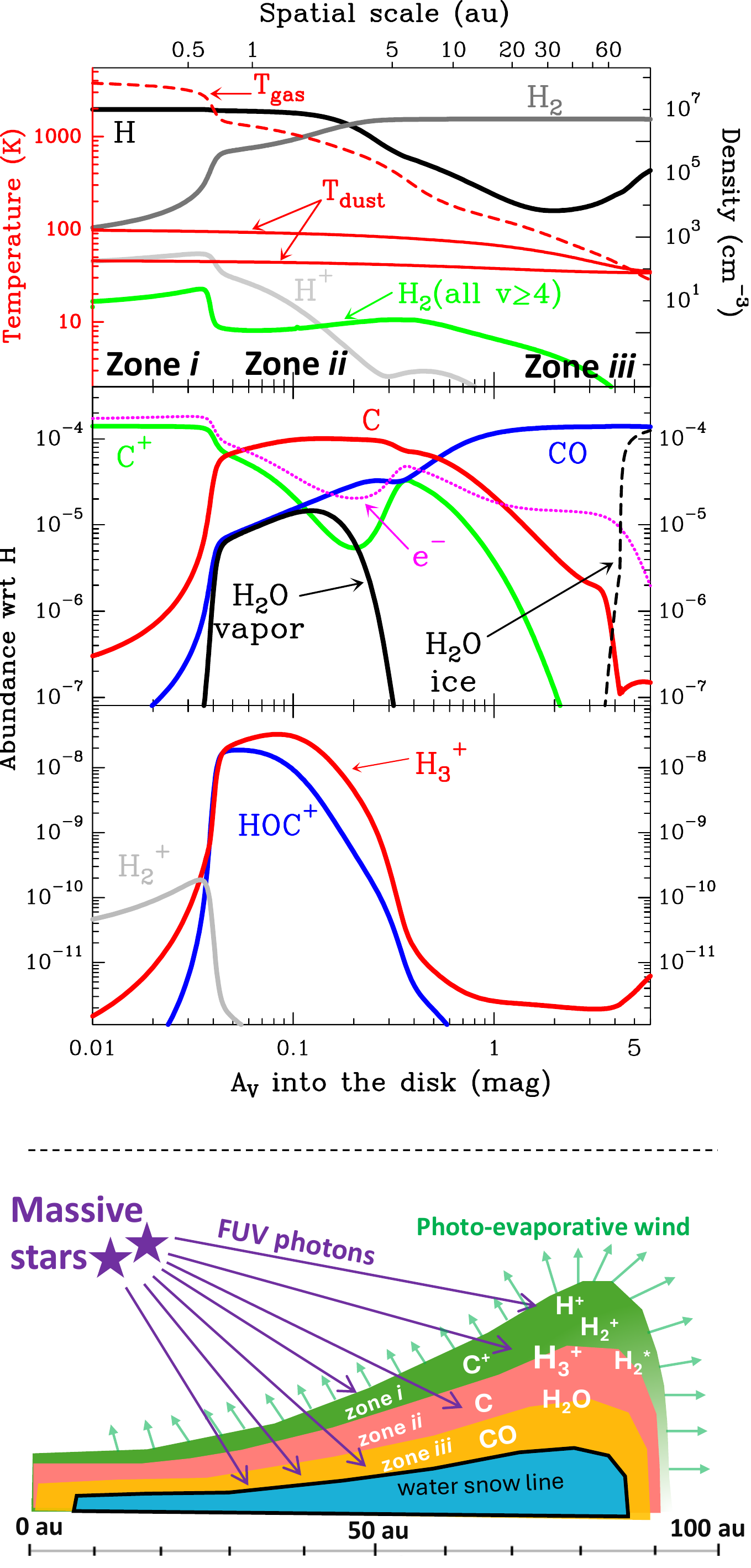}
\caption{Physical (approximately vertical) structure of 
\mbox{d203-506}.
\mbox{\textit{Upper panels}:} 
Model profiles of gas density, gas temperature, and $a^{-3.5}$ grain-size distribution, showing the highest and lowest dust temperatures (corresponding to $a_{\rm min}$ and $a_{\rm max}$ grain radii, respectively) as a function of depth into the PDR. 
 The green curve shows the density of H$_{2}^{*}$($v$\,$\geq$4). \textit{Lower panel}: Abundance profiles of selected species. \textit{Bottom sketch:}
 Simplified sketch illustrating the PDR component of an
externally irradiated protoplanetary disk. The different colored regions correspond to the chemical zones discussed in Sect.~\ref{sec:chemistry}.}  
\label{fig:chemical_model_d203-506}
\end{figure}

 \citet{Ilane25b}  reported the likely detection   of several  H$_{3}^{+}$  emission lines of the \( \nu_{2}=1 \rightarrow 0 \) band,
including the \mbox{$R$(1,0)}, \mbox{$Q$(1,0)},  \mbox{$Q$(5,$G$)}, \mbox{$R$(7,0)}, and \mbox{$R$(7,6)$^u$} lines\footnote{The NIR spectrum of this  disk is very rich and additional H$_{3}^{+}$ lines remain blended with lines of other species \citep{Ilane25b}.}.
The higher excitation lines are not detected in the ISM of galaxies
\citep{Pereira24} and suggest higher excitation conditions in \mbox{d203-506}.
Although clearly visible in the stacked spectrum, the reported lines are faint and close to the detection limit, leaving the presence of H$_3^+$ emission in \mbox{d203-506} open to interpretation.

 \section{Representative  model of the externally irradiated outer  disk and photoevaporative wind}\label{sec:PDR_reference}
 
 To guide our interpretation of the H$_3^+$ emission in externally \mbox{irradiated} disks, we first present a representative model of their PDR component, its structure, and  physical conditions. We assume that at the physical scales probed by JWST, the density distribution is smooth across the inner neutral wind and outer disk
 surface. Thus, as a first approximation, we adopted a constant density model.
We simulated the
disk PDR as a 1D stationary slab of gas and dust using an enhanced version of
the Meudon PDR\footnote{\url{https://pdr.obspm.fr/pdr_download.html}} code 
v7.0 \mbox{\citep{LePetit06}}. Since we were focusing on  \mbox{d203-506}, we adopted $G_0$\,=\,2$\times$10$^4$ and
\mbox{$\zeta$\,=\,10$^{-16}$\,s$^{-1}$} \citep{Berne24}. 
Our model solves the \mbox{$\lambda$-dependent} attenuation of FUV photons ($\lambda$\,$<$\,912\AA), considering dust extinction and line self-shielding. 
We treat the \mbox{$\lambda$-dependent} absorption and anisotropic scattering of FUV photons by dust grains that follow a uniform size distribution 
\mbox{$\propto$\,$a^{-3.5}$} \citep{Mathis77}, with \mbox{$a_{\rm min} = 0.02\,\mu$m} and 
\mbox{$a_{\rm max} = 1\,\mu$m} as the minimum and maximum grain radii
\citep[for details, see][]{Goico07}. This choice provides a total-to-selective extinction ratio of 
\mbox{$R_V \simeq 5.9$}
and a dust extinction cross section \mbox{$\sigma_{\rm ext}^{\rm d}$\,$=$\,$7 \times 10^{-22}$\,cm$^2$\,H$^{-1}$} at 1{,}000\,\AA\, (absorption plus scattering).
While these grains are bigger than ISM grains (where $\sigma_{\rm ext}^{\rm d}$ is $\sim$3 times larger), they reflect the modest grain growth expected in the upper layers of young irradiated disks \mbox{\citep[e.g.,][]{Stoerzer99,Birnstiel18}}.

Our detailed treatment of the penetration of FUV radiation, multilevel H$_2$ excitation
(up to \mbox{$v$\,=\,14}), thermal balance, and chemistry allows for the precise determination of the H/H$_2$ and C$^+$/C/CO transition layers, which is crucial for understanding the origin of the  H$_3^+$ emission. In order to account for the formation
of the outer water snow line, the model includes
gas-phase chemistry and simple grain surface chemistry for O, OH, H$_2$O, O$_2$, and CO
ice mantles \citep[see][]{Putaud19,Goicoechea21b}.
However, we do not model the different reactivities of \textit{ortho--} and \textit{para}--H$_2$ or H$_3^+$, nor do we compute the vibrational populations of chemical reaction products.
In addition, we neglect the wind dynamics 
\mbox{\citep[see e.g.,][]{Haworth20}}.

\begin{figure}[t]
\centering   
\includegraphics[scale=0.32, angle=0]{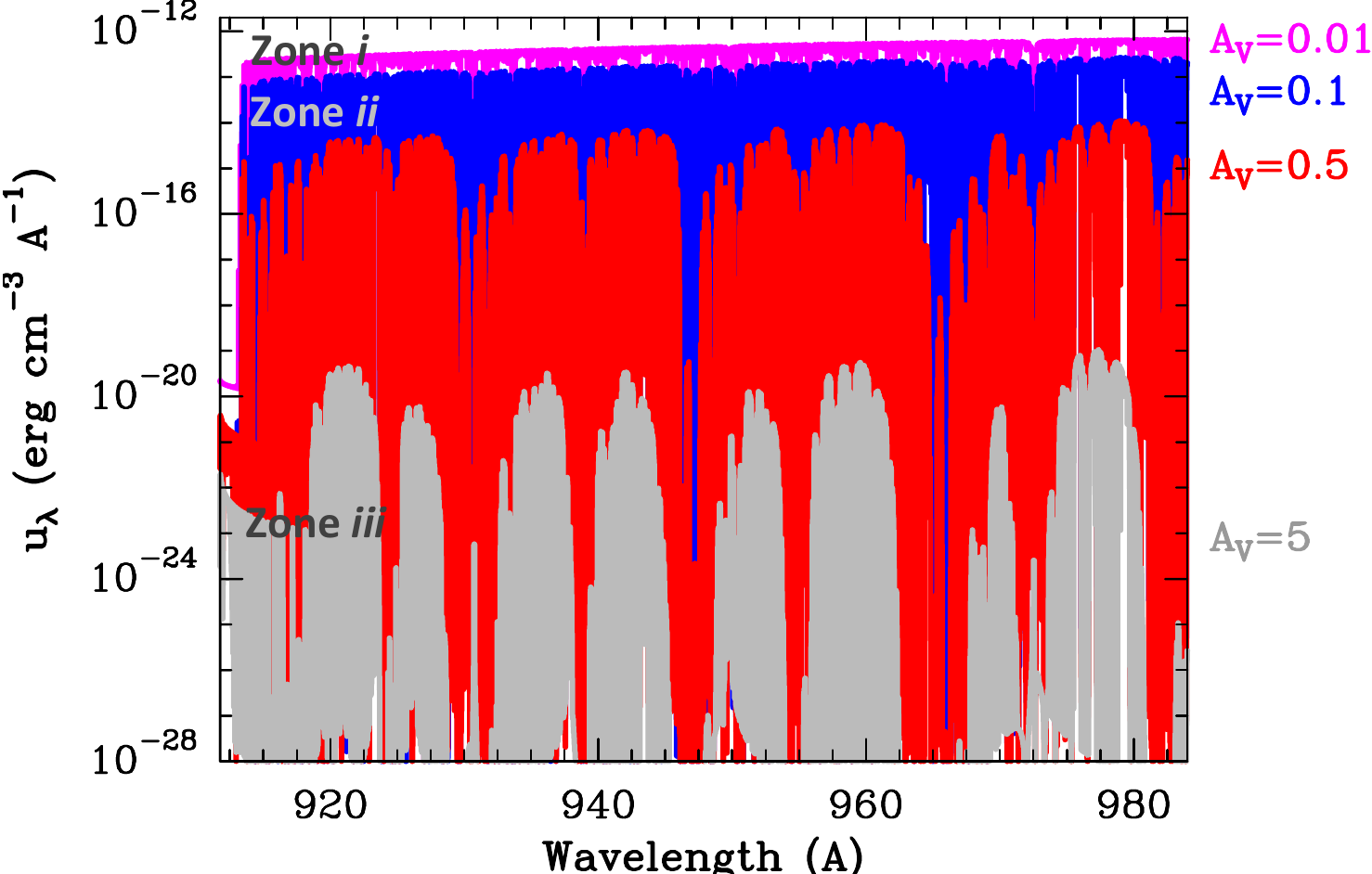}
\caption{Predicted $\lambda$-dependent FUV radiation field at different depth positions from the PDR surface into the disk. We show the local FUV energy density in units of erg\,cm$^{-3}$\,\AA$^{-1}$, from the Lyman cut to 12.6\,eV. }  
\label{fig:fuv_field_d203-506}
\end{figure}

Based on JWST studies of \mbox{d203-506}, we adopted a gas density \mbox{$n_{\rm H}$\,=\,$n$(H)\,+\,2$n$(H$_2$)\,=\,10$^7$\,cm$^{-3}$} for its PDR component
\citep[][]{Berne24,Goico24,Zannese24,Zannese25}. Figure~\ref{fig:chemical_model_d203-506}  dissects the  physical structure of the PDR  as a function of depth (along the external illumination direction and in magnitudes of visual extinction, $A_V$),  roughly representing the vertical structure of the outer disk and  wind. 
This figure   shows the decreasing gas temperature and increasing H$_2$ density  from the PDR edge to the more \mbox{FUV-shielded} disk interior. \mbox{Figure~\ref{fig:fuv_field_d203-506}} shows the FUV radiation field (in energy density) at different zones of the disk PDR
(in $A_V$). In this plot, we only show the high-energy FUV photon range from 13.6 to 12.6\,eV relevant to H$_2^*$ photoionization.
Owing to the low $G_0$/$n_{\rm H}$ ratio, the FUV spectrum is significantly blanketed by  H$_2$ absorption lines (\mbox{Fig.~\ref{fig:fuv_field_d203-506}}).

 H$_2$ molecules are photodissociated at the very surface of the PDR
 (\mbox{$A_V \lesssim 0.2$\,mag}). In these hot and predominantly atomic gas layers, the abundance of H atoms exceeds that of H$_2$. That is, \mbox{$x$(H)\,$>$\,$x$(H$_2$)} and
\mbox{$f$(H$_2$)\,=\,2$n$(H$_2$)\,/\,$n_{\rm H}$\,$<$\,2/3}, where $f$(H$_2$) is the molecular gas fraction. Owing to the high densities and temperatures, the excited rotational levels of \mbox{H$_2$($v = 0$)} become significantly populated
by inelastic collisions, with \mbox{$T_{\rm rot} \approx T_{\rm gas}$}.
Due to the de-excitation of \mbox{FUV-pumped H$_2$} electronic levels (Lyman and Werner bands), a significant fraction of the existing H$_2$ resides in excited vibrational states too \mbox{($v$\,$>$\,0}; see the upper panel of \mbox{Fig.~\ref{fig:chemical_model_d203-506}}).  Specifically, the predicted fractional abundance of H$_2^*$ molecules in vibrational states 
$v \geq 4$ (relative to those in the ground state) reaches 
$\sim$\,1\% (this fraction increases for lower gas densities, see specific models in  \mbox{Appendix~\ref{appendix-Orion Bar}}). 
A key feature of PDRs is that  the enhanced abundance of H$_2^*$ significantly 
impacts the chemistry
and the position\footnote{H$_2^*$ readily reacts with C$^+$ and O to form abundant CH$^+$ and OH 
\citep[as observed in \mbox{d203-506};][]{Zannese24,Zannese25}. These reactions are highly endoergic when H$_2$ ($v$\,=\,0, $J$\,=\,0), but they become fast at high $T_{\rm gas}$ and when H$_2^*$ is available, even shifting the \mbox{C$^+$/C} transition ahead of the \mbox{H/H$_2$} dissociation front (see \mbox{Fig.~\ref{fig:chemical_model_d203-506}}) and triggering the formation of specific molecular species in the PDR.} of the C$^+$/C/CO transition layers.
Owing to self-shielding and a reduced FUV flux, H$_2$ becomes more abundant than H at
 $A_V > 0.2$\,mag. This marks the so-called \mbox{``dissociation front''}, where 
$n$(H)\,$\simeq$\,$n$(H$_2$).
Most of the IR molecular emission detected by JWST in \mbox{d203-506} 
originates from these warm disk layers, where we predict 
\mbox{$T_{\rm gas}$\,$\approx$\,1{,}000\,K}, in agreement with the
observed H$_2$ rotational temperatures \citep{Berne24}.
The lowest panel of \mbox{Fig.~\ref{fig:chemical_model_d203-506}} shows a simplified sketch illustrating three  zones of the disk PDR (from the outside to the inside): predominantly atomic hot gas at the neutral wind edge (zone~$i$), 
 warm gas near the H/H$_2$ dissociation front (zone~$ii$), and colder, 
partially \mbox{FUV-shielded} disk gas (zone~$iii$). These zones also correspond to different regimes of the H$_3^+$ chemistry.

\begin{figure}[h]
\centering   
\includegraphics[scale=0.405, angle=0]{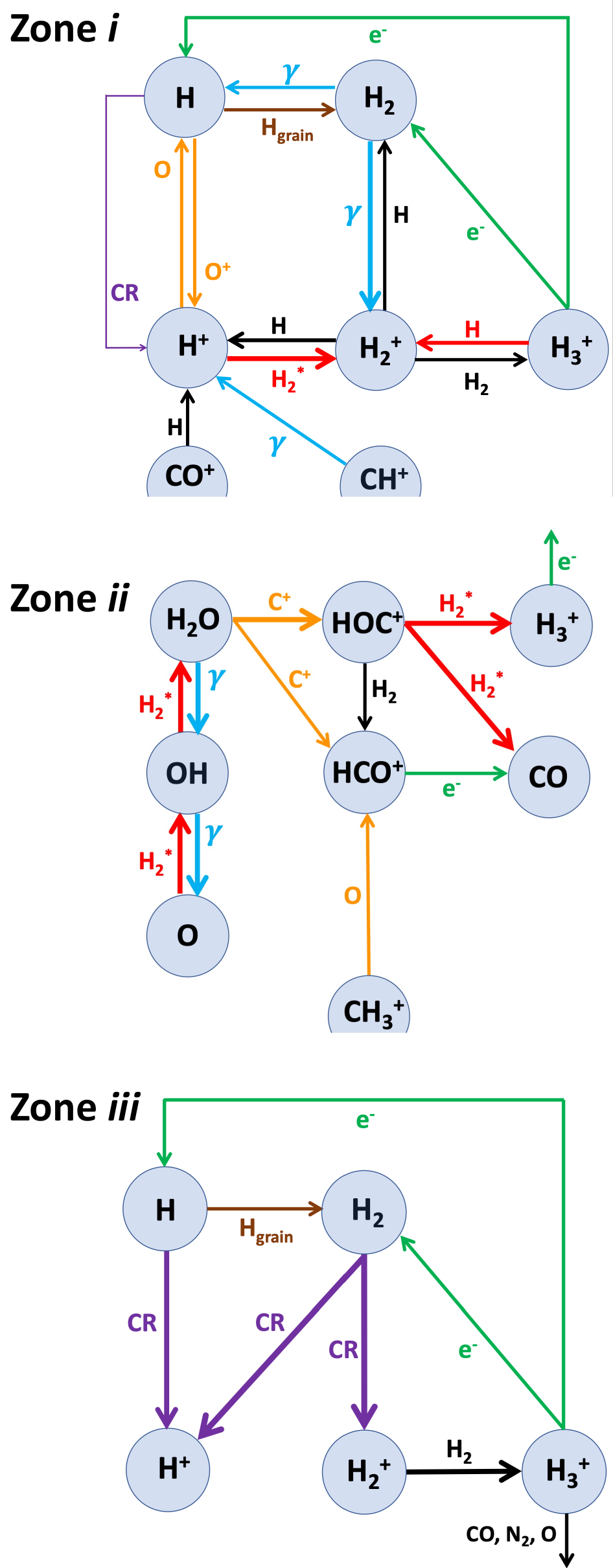}
\caption{Dominant H$_3^+$ formation and destruction pathways in  different zones of a  strongly irradiated protoplanetary disk (see \mbox{Fig.~\ref{fig:chemical_model_d203-506}}). 
 A thicker arrow represents a dominant
 chemical pathway. Red arrows show endoergic reactions. These reactions become fast at high $T_{\rm gas}$ and where significant H$_{2}^{*}$ exists. “CR” refers to ionization caused by cosmic rays and “$\gamma$” refers to FUV photons. Orange arrows represent ion-molecule and charge exchange reactions.
Green arrows represent dissociative recombinations with electrons. 
Incomplete circles indicate that these are not the starting points of the chemistries.}
\label{fig:Network_H3p}
\end{figure}

 \section{H$_3^+$ chemistry in three different disk regions}\label{sec:chemistry}

In this section, we introduce the main sources of H$_2$ ionization and the dominant pathways leading to H$_3^+$ in the three distinct zones of the disk PDR 
(\mbox{Fig.~\ref{fig:chemical_model_d203-506}}).
This analysis is based on the dominant H$_3^+$ formation and destruction mechanisms predicted by our reference model of \mbox{d203-506} across the disk PDR (see the chemical networks in \mbox{Figure~\ref{fig:Network_H3p}}).

\subsection{Zone i: Predominantly atomic PDR gas }
\label{subsec:component-i}

\begin{figure}[t]  
\centering    
\includegraphics[scale=0.4, angle=0]{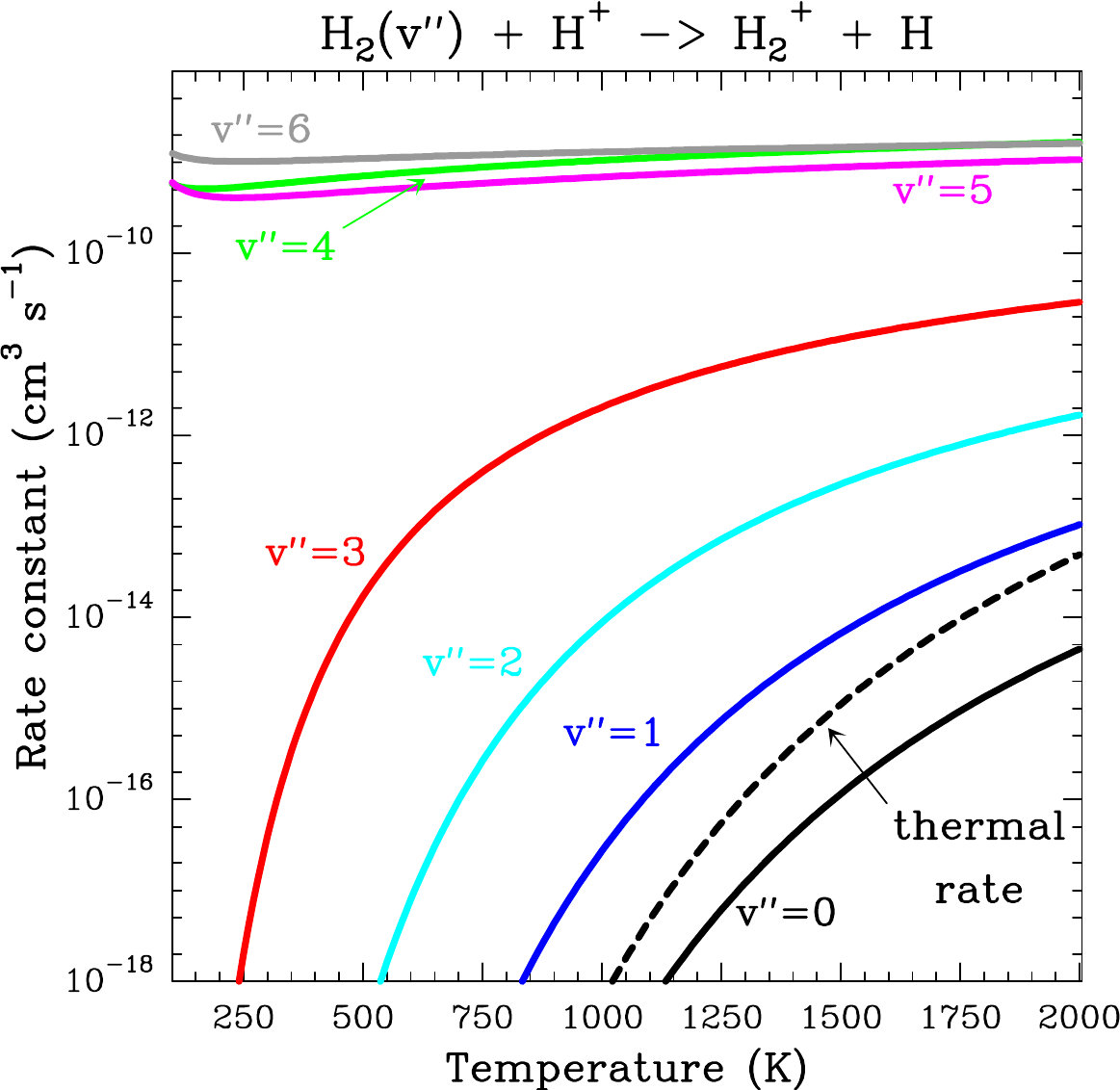}
\caption{Calculated vibrational-state-specific rate constants of \mbox{reaction}
\mbox{H$_2$($v$'')\,+\,H$^+$\,$\rightarrow$\,H$_2^+$\,+\,H} (reaction~\ref{reaction_hp}).  
The dashed curve represents the thermal\blue{$^6$} rate constant determined from thermal averages of these constants.}
\label{fig:New_rates_H2_Hp}
\end{figure}

This thin outermost zone ($A_V$\,$<$\,0.03\,mag, less than one astronomical unit, au) 
is exposed to the nearly unshielded external FUV field and is characterized by very high temperatures
($\simeq$\,3{,}000\,K in this model), fairly low $f$(H$_2$), and significant
relative abundances of H$_2^*$.
The dominant H$_2$ ionization pathway is the reaction 
\begin{equation}
{\rm H_2}(v'')\,+\,{\rm H^+}\,\rightarrow\,{\rm H_2^+}(v')\,+\,{\rm H},
\label{reaction_hp}
\end{equation}
where $v''$ and $v'$ denote the vibrational levels of the reactant and product molecules, respectively, in their ground electronic state.
This process -- both charge transfer (CT) and reactive charge transfer (RCT) -- is very endoergic, by about 1.83\,eV\,$\simeq $\,21,200\,K, when \mbox{$v''$\,=\,0}  
\citep[][]{Ichiara00,Krstic02,Plasil12,Sanz-Sanz21}. Thus, it is a negligible process in cold molecular gas.
However, it becomes accessible at high temperatures \mbox{($T_{\rm gas}>1{,}000$\,K)} and as higher
$v''$ levels of  H$_2^*$ are populated by  \mbox{de-excitation} of \mbox{FUV-pumped} H$_2$. In Appendix~\ref{App:Charge_Tramsfer}  we describe the quantum dynamical  methods we used to calculate accurate \mbox{vibrational-state} dependent  rate constants for 
\mbox{reaction~(\ref{reaction_hp})}.
\mbox{Table~\ref{table:rates}} summarizes the derived 
$v''$-state specific rate constants as well as the thermal\footnote{To  distinguish between the role of high gas temperatures  and the effect of \mbox{nonthermal}  populations of FUV-pumped H$_2^*$($v''$) on the chemistry, it is useful to determine the thermal rate constants. These are
  \mbox{Boltzmann} averages of the individual $v''$-state-specific rate constants, defined as
  \begin{equation}
  k_{\rm th}(T) =  \frac{\sum_{v''=0}^{14}\,k_{v''}(T) \,e^{-E_{v''}/k_{\rm B}T} }
                        {\sum_{v''=0}^{14}\,e^{-E_{v''}/k_{\rm B}T}}.  
  \label{eqn:thermal_rate}
\end{equation}

Figure~\ref{fig:New_rates_H2_Hp} shows the resulting thermal rates 
for reaction~(\ref{reaction_hp}) 
 as a dashed black curve.
 At low temperatures \mbox{($T_{\rm gas}$\,$\ll$\,$E_{v''}\,({\rm H_2})/k_{B}$)}, 
 as in most applications involving cold molecular gas, the thermal rate  is roughly that of \mbox{H$_2$($v''$\,=\,0)}. That is,
\mbox{$k_{\rm th}(T)$\,$\simeq$\,$k_{v''=0}(T)$} \citep[e.g.,][]{Agundez10}.
In the case of \mbox{reaction~(\ref{reaction_hp})}, $k_{\rm th}$ is significantly larger than $k_{v''=0}$ because the rate constants increase by orders of magnitude when H$_2$ is vibrationally excited. For reaction~(\ref{reaction_hocp}), the differences between $k_{v''=0}$ and $k_{v''\geq1}$
 are not that large, and $k_{\rm th}$ becomes slightly higher than $k_{v''=0}$ at high temperatures, where the vibrationally excited levels start to be populated.} rate, derived from these constants,
in the form of Arrhenius-like fits.
\begin{table}[t]
  \begin{center}
    \caption{\label{table:rates} Rate constants discussed in this work.}
  \vspace{-0.5cm}
 \resizebox{\linewidth}{!}{
  \begin{tabularx}{\linewidth}{l c c c @{\vrule height 10pt depth 5pt width 0pt}}
      \hline \hline
      Reaction & $\alpha$            & $\beta$ & $\gamma$\\
               &  (cm$^3$\,s$^{-1}$) &         &  (K)\\      
      \hline
      H$_2$($v$=0)\,+\,H$^+$\,\,$\rightarrow$\,\,H$_2^+$\,+\,H & 2.70e$-$10 &  & 22000  \\
      H$_2$($v$=1)\,+\,H$^+$\,\,$\rightarrow$\,\,H$_2^+$\,+\,H & 4.00e$-$10 &  & 16500  \\        
      H$_2$($v$=2)\,+\,H$^+$\,\,$\rightarrow$\,\,H$_2^+$\,+\,H & 3.20e$-$10 &  & 10500  \\
      H$_2$($v$=3)\,+\,H$^+$\,\,$\rightarrow$\,\,H$_2^+$\,+\,H & 5.74e$-$11 & $+$0.76 & 4254  \\
      H$_2$($v$=4)\,+\,H$^+$\,\,$\rightarrow$\,\,H$_2^+$\,+\,H & 3.74e$-$10 & $+$0.75 & $-$126  \\
      H$_2$($v$=5)\,+\,H$^+$\,\,$\rightarrow$\,\,H$_2^+$\,+\,H & 2.28e$-$10 & $+$0.77 & $-$182  \\
      H$_2$($v$=6)\,+\,H$^+$\,\,$\rightarrow$\,\,H$_2^+$\,+\,H & 7.58e$-$10 & $+$0.37 & $-$92   \\
      H$_2$(thermal)\,+\,H$^+$\,\,$\rightarrow$\,\,H$_2^+$\,+\,H & 1.32e$-$09 & $+$0.42 & 21993  \\
      \hline
      H$_2^+$\,+\,H\,\,$\rightarrow$\,\,H$_2$\,+\,H$^+$ &   6.64e$-$10 &  $+$0.21  &  $-$2.0 \\
      \hline
      HOC$^+$\,+\,H$_2$($v$=0)\,$\rightarrow$\,H$_3^+$\,+\,CO   & 1.25e$-$11 & $+$1.07 & $-$80.5 \\
      HOC$^+$\,+\,H$_2$($v$=1)\,$\rightarrow$\,H$_3^+$\,+\,CO   & 3.44e$-$10 & $+$0.06 &    3.4\\
      HOC$^+$\,+\,H$_2$($v$=2)\,$\rightarrow$\,H$_3^+$\,+\,CO   & 4.77e$-$10 & $-$0.01 &    7.1 \\
      HOC$^+$\,+\,H$_2$($v$=3)\,$\rightarrow$\,H$_3^+$\,+\,CO   & 5.34e$-$10 & $-$0.01 & $-$1.9 \\
      HOC$^+$\,+\,H$_2$($v$=4)\,$\rightarrow$\,H$_3^+$\,+\,CO   & 5.40e$-$10 & $+$0.01 & $-$12.2 \\
      \hline
      HOC$^+$\,+\,H$_2$($v$=0)\,$\rightarrow$\,HCO$^+$\,+\,H$_2$   & 2.37e$-$10 & $-$0.37 &  20.0\\
      HOC$^+$\,+\,H$_2$($v$=1)\,$\rightarrow$\,HCO$^+$\,+\,H$_2$   & 2.16e$-$10 & $-$0.35 &  20.0\\
      HOC$^+$\,+\,H$_2$($v$=2)\,$\rightarrow$\,HCO$^+$\,+\,H$_2$   & 1.66e$-$10 & $-$0.33 &  20.0\\
      HOC$^+$\,+\,H$_2$($v$=3)\,$\rightarrow$\,HCO$^+$\,+\,H$_2$   & 1.47e$-$10 & $-$0.37 &  20.0\\
      HOC$^+$\,+\,H$_2$($v$=4)\,$\rightarrow$\,HCO$^+$\,+\,H$_2$   & 1.19e$-$10 & $-$0.29 &  20.0\\
      \hline
      HOC$^+$\,+\,H\,$\rightarrow$\,CO$^+$\,+\,H$_2$              & 5.10e$-$10 &  & 7125\\
      HCO$^+$\,+\,H\,$\rightarrow$\,CO$^+$\,+\,H$_2$              & 1.30e$-$09 &  & 24500\\
      H$_3^+$\,+\,H\,$\rightarrow$\,H$_2^+$\,+\,H$_2$             & 2.10e$-$09 &  & 20000\\
      
      \hline
      \end{tabularx}}
      
      \tablefoot{Rate constants from fitting the Arrhenius-like form
     $k\,(T)$\,=\,$\alpha\,(T/300\,{\rm K})^\beta \,{\rm exp}(-\gamma/T)$
     to reaction rates computed in this study, and valid in the \mbox{$T$\,$\simeq$\,100\,--\,2{,}000\,K} range. For the last three endoergic reactions, see footnote \blue{10}.}
      
  \end{center}
\end{table}
For H$_2$($v''$) in low vibrational states ($v''$\,$<$\,4) and at warm to cold gas temperatures ($T_{\rm gas}<$\,1{,}000\,K), the reaction~(\ref{reaction_hp})  is extremely slow
(see \mbox{Fig.~\ref{fig:New_rates_H2_Hp}}). However, for \mbox{$v''$\,$\geq$\,4} and for 
$T_{\rm gas}>$\,1{,}000\,K, the CT reaction becomes energetically accessible and the reaction rate much faster.

Given that H$_3^+$ formation via cosmic-ray ionization of H$_2$ is negligible in this zone, the central question becomes how hydrogen atoms are ionized.
A  fraction of H atoms are directly  ionized by cosmic rays. In addition, the photodissociation of CH$^+$ and reactions of CO$^+$ with H atoms become  significant \mbox{``chemical sources''} of H$^+$ in high-$G_0$ environments. These reactive molecular ions\footnote{Reactive ions such as H$_2^+$, CH$^+$, and HOC$^+$ are transient molecules for which the timescale of \mbox{reactive} collisions (leading to molecular destruction) is comparable to or shorter than that of inelastic collisions \citep[e.g.,][]{Black98}.
Properly treating their molecular excitation usually requires incorporating chemical formation and destruction rates into the statistical equilibrium level equations
(see \mbox{Appendix~\ref{appendix-grosbeta}}).}
  are  natural products of PDR chemistry \mbox{\citep[e.g.,][]{Sternberg95}} and readily form 
as $T_{\rm gas}$ and the H$_2^*$ abundance increase in the disk PDR.

In this zone, H$^+$  ions are destroyed by reactions with
O atoms, 
\begin{equation}
\rm{H^+\,+\,O(^3P_J)\,\rightarrow\,H\,+\,O^+}.
\label{reaction_cto}
\end{equation}
This  resonant CT is 
endoergic\footnote{We predict the following O($^3$P$_J$) fine-structure level
populations in zones~$i$ and $ii$: $\sim$60\,\% in $J=2$, $\sim$30\,\% in $J=1$, and $\sim$10\,\% in $J=0$. In zone~$iii$, most of the
oxygen is in the ground fine-structure level.} 
when the oxygen atom is the ground  fine-structure state ($J$\,=\,2) \citep[][]{Chambaud80,Stancil99,Spirko03}, with a rate constant 
\mbox{$\propto$\,exp($-$230\,K\,/\,$T_{\rm gas}$)}. Thus, it is much faster at high $T_{\rm gas}$, so that
H$^+$ ions follow \mbox{reaction~(\ref{reaction_cto})} instead of recombining with
electrons.  
Still, a small fraction of H$^+$ ions forms H$_2^+$ through \mbox{reaction~(\ref{reaction_hp})}.
The inverse CT of reaction~(\ref{reaction_cto}) \mbox{(H\,+\,O$^+$)} is exothermic and quickly returns
most of the charge to H$^+$.

Although FUV photons (below the Lyman limit) cannot ionize molecular hydrogen when H$_2$ is in its
lowest vibrational states \citep[as \mbox{IP(H$_2$)\,=\,15.43\,eV or 
124,417.491\,cm$^{-1}$};][]{Liu09},  H$_2^*$ molecules  in \mbox{$v''$\,$\geq$\,4}  can be ionized\footnote{The \mbox{H$_2$($v$=4, $J$=0)} level is at an energy of
1.89\,eV (or 21504.7\,cm$^{-1}$) above the ground state \citep[e.g.,][]{Komasa11}. Thus,  FUV photons with  \mbox{$E_{\lambda}$\,$=$\,IP(H$_2$)\,--\,$E_{\rm H_2(\textit{v}\geq4)}<$\,13.6\,eV} produce ionization and lead to H$_2^+$.} by 
FUV radiation. That is,
\begin{equation}
{\rm H_2}(v''\,\geq\,4)\,+\,{\rm FUV}\,\rightarrow\,{\rm H_2^+}(v')\,+\,{\rm e^-}.
\label{reaction_photoi}
\end{equation}
This is a FUV-driven source of
H$_2$ ionization,
and of  electrons, that has  not been considered in PDR models before. This mechanism is also
a destruction process of H$_2$ levels with $v''\,\geq\,4$. 
We determined the role of H$_2^*$ photoionization by considering the \mbox{state-to-state} photoionization 
 cross sections, \mbox{$\sigma_{\lambda}$\,($v''$)\,=\,$\sum_{v'}$\,$\sigma_{\lambda}$($v''$$\rightarrow$$v'$)},  computed by \cite{Ford75}. These cross sections are on the order of 
 \mbox{$\sim$10$^{-18}$\,cm$^{-2}$}, and require FUV photons
with energies above \,12.6\,eV ($\lambda$\,$<$\,984\,\AA, see \mbox{Fig.~\ref{fig:fuv_field_d203-506}}).
\mbox{Appendix~\ref{appendix-photo-x-sections}} describes the details of our calculations.

The photoionization rate of H$_2^*$ from a given $v''\geq4$ 
level, \mbox{$\kappa_{\rm phi}$(H$_2$,\,$v''$)}, is given by
\begin{equation}
\kappa_{\rm phi}({\rm{H_2}},\,v'')
= \frac{1}{h} \int_{912}^{\lambda_t} \sigma_{\lambda}(v'')\, u_{\lambda}\,\lambda\, d\lambda\,\,\,\,\,\,\,\,{\rm{(s^{-1})}},
\label{reaction_kappa_phi} 
\end{equation}
where $u_{\lambda}$  is the local energy density of the FUV radiation field (in \mbox{erg\,cm$^{-3}$\,\AA$^{-1}$})
at a given disk position (see \mbox{Fig.~\ref{fig:fuv_field_d203-506}}). Hence, the total H$_2^+$ formation rate due to 
\mbox{H$_2^*$($v''$$\geq$4)} photoionization, \mbox{$F_{\rm phi}$(H$_2^+$)}, is given by
\begin{equation}
F_{\rm phi}({\rm H_2^+}) = \sum_{v''} n({\rm H_2},\,v'')\,\kappa_{\rm phi}({\rm{H_2}},\,v'')\,\,\,\,\,\,\,\,{\rm{(cm^{-3}\,s^{-1})}},
\label{reaction_F_phi}
\end{equation}
where $n({\rm H_2},\,v'')$ (in cm$^{-3}$) are the local H$_2^*$($v''$)  level populations.

In zones $i$ and $ii$, reaction~(\ref{reaction_h3p}) is not the dominant destruction pathway of H$_2^+$. Instead, owing to the high abundance of H atoms,
the reverse process of  
\mbox{reaction~(\ref{reaction_hp})},
\begin{equation}
{\rm H_2^+}(v'')\,+\,{\rm H}\,\rightarrow\,{\rm H_2}(v')\,+\,{\rm H^+},
\label{reaction_hpr}
\end{equation}
is the dominant destruction pathway for H$_2^+$.
This reaction is very exoergic, and has been investigated in the laboratory 
\citep[][]{Karpas79} and by quantum dynamical methods \mbox{\citep[][]{Sanz-Sanz21}}.
In Appendix~\ref{App:Charge_Tramsfer} we summarize how we determined the state-dependent  rate constants for this reaction (CT and RCT) in the temperature range
\mbox{$T_{\rm gas}$\,$\simeq$\,100\,--\,2{,}000\,K}.
\mbox{Table~\ref{table:rates}} shows the computed thermal rate 
constant -- independent of the vibrational state of \ch{H2+}
(as we do not explicitly follow the excitation of \ch{H2+} molecules) --
that we used in our models.
Reactions of H$_2^+$ with H limit the formation of H$_3^+$ in \mbox{FUV-irradiated} gas with relatively low $f$(H$_2$).
In general, reactions involving H atoms can constitute an efficient destruction pathway for molecules at high temperatures because these processes are typically highly 
endoergic\footnote{Our chemical network includes endoergic reactions of H atoms with H$_3^+$, HCO$^+$, HOC$^+$ 
\mbox{(Table~\ref{table:rates})}, and other molecules. If the forward rate is not known, we adopt the rate constant of the reverse reaction multiplied by $e^{-\Delta E/T}$, where $\Delta E$ is the endothermicity of the forward reaction \citep[see e.g.,][]{PineaudesForets86}.}. However, they only impact a thin layer at the PDR edge, where temperatures are very high and \mbox{$x$(H)\,$\gg$\,$x$(H$_2$)}.

\subsection{Zone ii: Molecular PDR gas}
\label{subsec:component-ii}

Figure~\ref{fig:Network_H3p} (middle panel)  shows the  gas-phase reactions leading to H$_3^+$ in zone~$ii$, near the dissociation front  (at $A_V$\,$\simeq$\,0.2\,mag in this reference model; Fig.~\ref{fig:chemical_model_d203-506}).
Because of its higher gas column density ($0.03 < A_V < 0.3$\,mag, corresponding to a vertical scale of \mbox{$\sim 5$--$10$\,au}) and elevated temperatures ($T_{\rm gas} \sim 1{,}000$\,K), zone~$ii$ is predicted to be the primary contributor to the observed H$_3^+$ column density in d203-506.
 \mbox{Remarkably}, the chemistry responsible for H$_3^+$ formation in this  zone is largely independent of cosmic-ray ionization, but ultimately related with the presence of 
enhanced abundances of C$^+$  and  water vapor in the outer disk surface.
We find that H$_3^+$ forms mainly through reaction 
\begin{equation}
{\rm H_2}(v'')\,+\,{\rm  HOC^+}\,\rightarrow\,{\rm H_3^+}\,+\,{\rm CO},
\label{reaction_hocp}
\end{equation}
 which is endoergic by $\sim$1{,}500 K when $v''$\,=\,0 (including zero-point energies, ZPEs, see Appendix~\ref{appendix-hocp_h2} and  \citet{Klippenstein10}). These products have been observed in experiments at room temperature
 \citep[e.g.,][]{Freeman87}. However, there is no evidence of H$_3^+$ formation at low temperatures \citep[at least at 25\,K; see][]{Smith02}.

\begin{figure}[t]  
\centering    
\includegraphics[scale=0.395, angle=0]{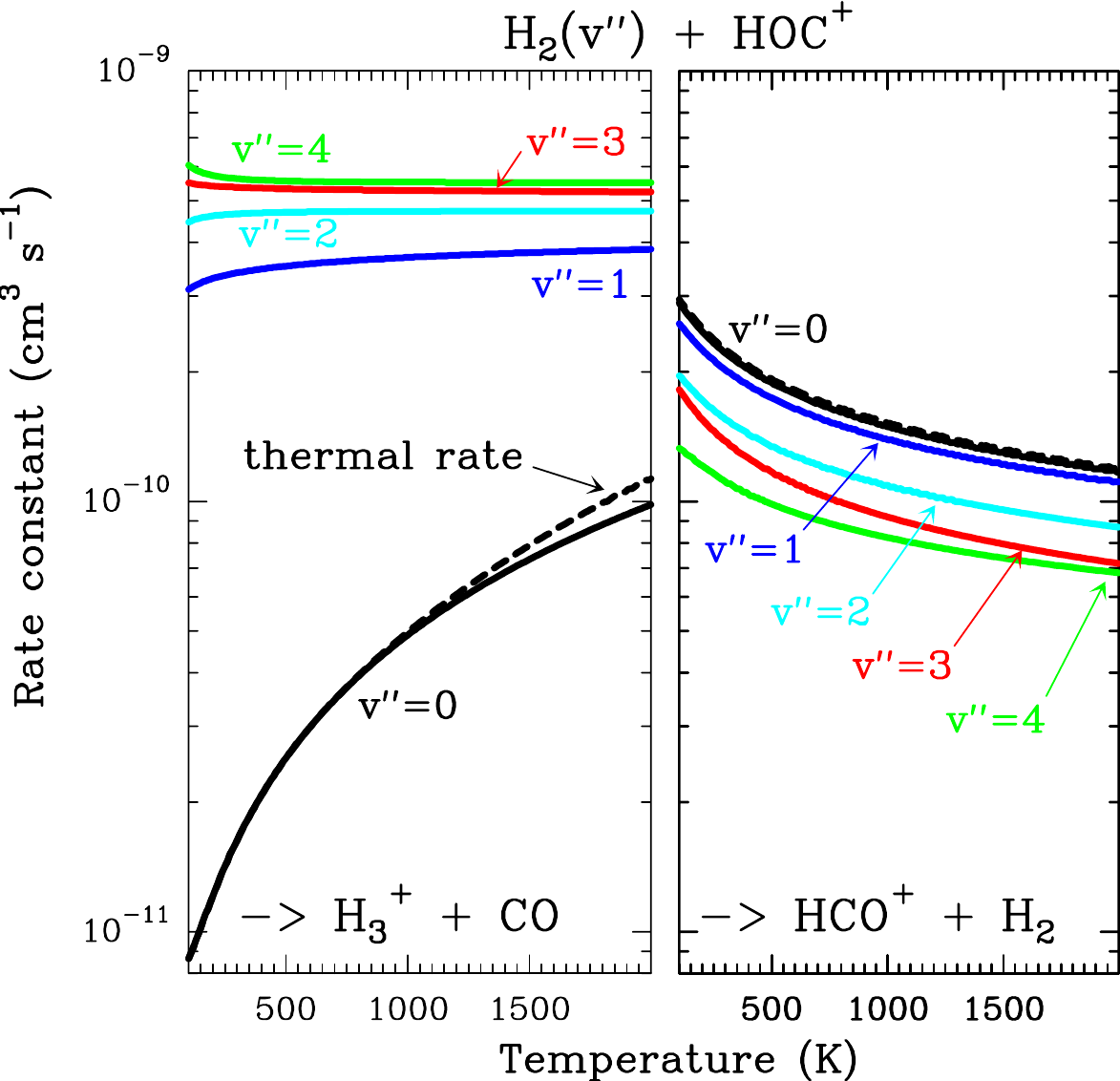}
\caption{Calculated H$_2$ vibrational-state-specific rate constants of \mbox{reaction}
\mbox{H$_2$($v$'')\,+\,HOC$^+$} producing \mbox{H$_3^+$\,+\,CO}  (left box)
and \mbox{HCO$^+$\,+\,H$_2$}  (right box).
The dashed curves 
represent the thermal\blue{$^6$} rate constant determined from thermal averages of the
state-dependent rates.}
\label{fig:New_rates_HOCp}
\end{figure}

  Reaction~(\ref{reaction_hocp})
competes with the isomerization reaction
\begin{equation}
{\rm H_2}(v'')\,+\,{\rm  HOC^+}\,\rightarrow\,{\rm HCO^+}\,+\,{\rm H_2},
\label{reaction_hocp-iso}
\end{equation}
which is exoergic \citep[][]{Smith02}. 
In \mbox{Appendix~\ref{appendix-hocp_h2}}
we describe the ZPE-corrected quasi-classical trajectory dynamical methods we used to calculate the \mbox{$v''$-state-dependent} rate constants of reactions  (\ref{reaction_hocp}) and (\ref{reaction_hocp-iso}).
Figure~\ref{fig:New_rates_HOCp}  show the resulting  rates. We find
that for \mbox{H$_2$($v''$\,=\,0,\,$J''$\,=\,0)}   and for low gas temperatures
($T_{\rm gas}$\,$<$100\,K), the isomerization reaction~(\ref{reaction_hocp-iso}) is the relevant  channel. However, for H$_2^*$, and as $T_{\rm gas}$ increases, the  channel leading to \mbox{H$_3^+$\,+\,CO} becomes increasingly 
relevant\footnote{Reaction~(\ref{reaction_hocp}) becomes exoergic for H$_2^*$
in $v''=0$ and $J''\geq4$.}, and becomes 
the primary pathway for H$_3^+$ formation  in this disk zone.

HOC$^+$  is the less stable isomer of the
widespread  HCO$^+$ cation. HOC$^+$ is a 
reactive molecular ion readily seen
in interstellar PDRs  \citep[e.g.,][]{Ziurys95,Fuente03,Liszt04,Savage04,Goico09,Goico17}. Its main formation pathway is the very exoergic reaction
\begin{equation}
\rm{C^+\,+\,H_2O\,\rightarrow\,HOC^+/HCO^+\,+\,H}.
\label{reaction_cp_h2o}
\end{equation}
This reaction has been studied in the laboratory \citep{Martinez08,Yang21},
implying a branching ratio to HOC$^+$ of $\gtrsim$\,0.68 because
the formation of HCO$^+$ requires carbon insertion into the \mbox{O--H} bond, which is less likely \citep[e.g.,][]{Ishikawa01}. Experimental studies suggest a reaction rate of $\gtrsim$\,10$^{-9}$\,cm$^3$\,s$^{-1}$ and a slight inverse temperature dependence \citep{Martinez08,Yang21}.
Therefore, the formation of abundant HOC$^+$ (and thus H$_3^+$) in the disk's PDR is directly linked to the presence of FUV photons, which can ionize C atoms, and to the existence of abundant water vapor, enabled by elevated gas temperatures.

\mbox{Figure~\ref{fig:Network_H3p}} (middle panel) shows the endoergic gas-phase reactions that lead to the formation of H$_2$O at warm temperatures.
The initial step involves the formation of the OH radical, 
\begin{equation}
{\rm O}\,+\, {\rm H_2}(v'')\,\rightarrow\,{\rm OH}(v')\,+\,{\rm H}.
\label{reaction_oh}
\end{equation}
This reaction is endoergic by $E$/$k$\,$\sim$\,770\,K and has an activation barrier of 
\mbox{$E_{\rm a}$/$k$\,$\sim$\,6900\,K}  from \mbox{H$_2$($v$=0)}
\citep[][]{Veselinova21}. The second step is reaction
\begin{equation}
{\rm OH}\,+\, {\rm H_2}(v'')\,\rightarrow\,{\rm H_2O}\,+\,{\rm H},
\label{reaction_h2o}
\end{equation}
which is exoergic, but  has a barrier\footnote{Since state-dependent rate constants,
$k_{v,J}$($T_{\rm gas}$), do not exist for \mbox{reaction~(\ref{reaction_h2o})},
we modeled them by adopting state-dependent rate constants where the energy $E_{v,J}$ of each H$_{2}^{*}$ ro-vibrational state is
\mbox{subtracted} from the reaction barrier $\Delta E$ (when \mbox{$\Delta E > E_{v,J}$}).} of a few \mbox{thousand} Kelvin. Thus, high
$T_{\rm gas}$ conditions, and the presence of H$_2^*$,  lead to water vapor in this PDR zone.
All in all, the strong external FUV radiation  initiates a cycle of H$_2$O formation, photodissociation, and reformation. This cycle has been observationally confirmed  by the detection of rotationally hot OH -- the signature of H$_2$O photodissociation -- and of vibrationally excited OH -- the signature of reaction~(\ref{reaction_oh}) -- in  \mbox{d203-506} \citep{Zannese24}.

\begin{figure*}[ht]
\centering   
\includegraphics[scale=0.415, angle=0]{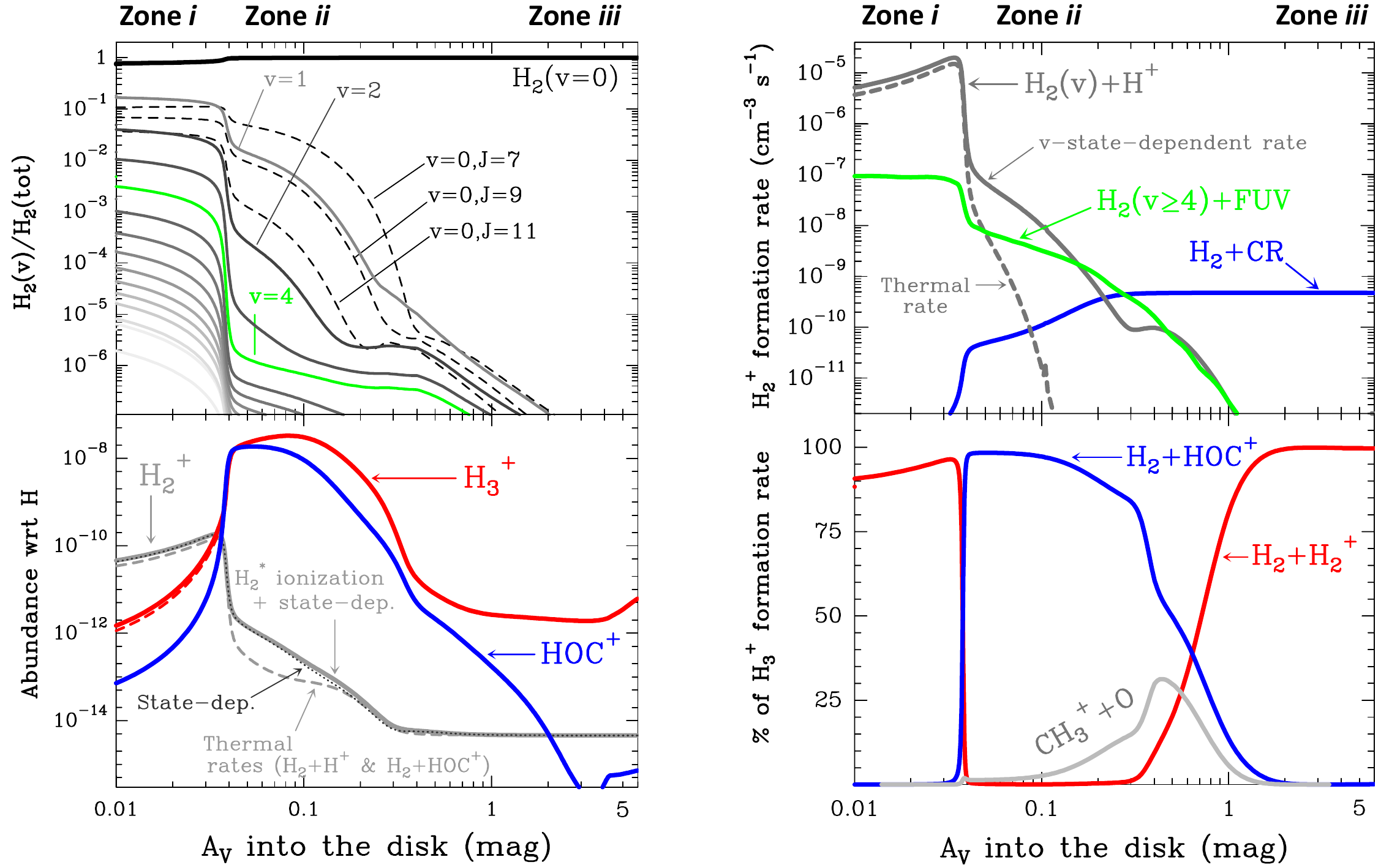}
\caption{Models  of the PDR component in \mbox{d203-506}. \textit{Left panels}:
H$_2$ relative populations and abundances.
The upper panel shows the normalized distribution of H$_2$($v$) populations across different vibrational levels (solid curves), along with the rotationally excited H$_2$($v$\,=\,0, $J$\,=\,7, 9, 11) levels (dashed curves).  The lower panel displays abundance profiles of H$_2^+$, H$_3^+$, and HOC$^+$.
Continuous curves show results for models using state-dependent rate constants for reactions~(\ref{reaction_hp}), (\ref{reaction_hocp}), (\ref{reaction_hocp-iso}), and for H$_2$($v$$\geq$4) photoionization (the complete reference model).
   Dotted curves are for models neglecting H$_2$ photoionization. 
 Dashed curves are for models neglecting H$_2$ photoionization  and using the thermal rate constants 
 for reactions (\ref{reaction_hp}),  (\ref{reaction_hocp}), and  (\ref{reaction_hocp-iso}).
\mbox{\textit{Right panels:}} Chemical formation rates. H$_2^+$ formation rates as a function of depth into the PDR (upper box). Contribution (in \%)  to the 
H$_3^+$ formation rate as a function of depth into the PDR (lower box).
}  
\label{fig:chemical_rates_H3p_d203-506}
\end{figure*}

 \subsection{Zone iii: Partially FUV-shielded molecular gas}
\label{subsec:component-iii}

Figure~\ref{fig:Network_H3p} (lower panel) shows 
the dominant gas-phase reactions leading to H$_3^+$ in FUV-shielded (or low illumination) molecular gas, where \mbox{$A_V$\,$\gtrsim$\,5\,mag}. 
The initial step involves cosmic-ray ionization of H$_2$:
\begin{equation}
{\rm H_2}\,+\, {\rm CR}\,\rightarrow\,{\rm H_2^+}\,+\,{\rm e^{-}}\,+\,{\rm CR'}.
\label{reaction_cr}
\end{equation}
This is the \mbox{``standard''} chemistry of interstellar clouds with \mbox{$f$(H$_2$)\,$\rightarrow$\,1} \citep[e.g.,][]{Dalgarno06}.
 In steady state, the density of H$_3^+$ molecules is simply given by
\begin{equation}
n({\rm H_3^+}) \simeq \frac{\zeta\,n({\rm H_2})}{\alpha_{\rm e}\,n({\rm e})\,+\, \sum_{\rm X} k_x\,n(\rm  X)},
\end{equation}
where we consider H$_3^+$ formation through reaction~(\ref{reaction_h3p}) and H$_3^+$ destruction through dissociative recombinations with electrons (with a rate constant \mbox{$\alpha_{\rm e}$\,$\sim$\,$T^{-0.5}$}) and proton transfer reactions with X\,=\,CO, N$_2$, O, and so on, 
with a rate constant, $k_{\rm x}$.
We note that this expression is independent of the rate constant
of \mbox{reaction~(\ref{reaction_h3p})}. In \mbox{FUV-shielded} dense molecular gas, $n$(e) is typically low,
\mbox{$\alpha_{\rm e}\,n({\rm e})$\,$\ll$\,$\sum_{\rm x} k_x\,n(\rm  x)$}, and 
\mbox{$x$(H$_3^+$)} is proportional to $\zeta$. 
This regime dominates deeper inside the cold midplane ($T_{\rm gas} < 100$\,K), but results in lower local H$_3^+$ abundances--$x$(H$_3^+$) of several $10^{-11}$  for $\zeta = 10^{-16}$\,s$^{-1}$ (or several $10^{-12}$ if  $\zeta =10^{-17}$\,s$^{-1}$) -- compared to the PDR, where  H$_3^+$ abundances are typically several $10^{-8}$.

\section{Discussion} 
\label{sec:discussion}

Here, we analyze and discuss our H$_3^+$ photochemical and excitation models, along with the sensitivity of the results to variations in the external radiation field
and the cosmic-ray ionization rate.

 \subsection{Photochemical model results}
\label{sec:chemical_models}

\mbox{Figure~\ref{fig:chemical_rates_H3p_d203-506}} (left panel, upper box) shows the predicted distribution of H$_2$ abundances (relative to the total H$_2$ density) across various rotationally and vibrationally excited H$_2$ levels as a function of depth into the disk.
The lower box shows the resulting 
H$_2^+$, H$_3^+$, and HOC$^+$ abundance profiles. 
In this box, continuous curves show the results from models that use the specific state-dependent rate constants for \mbox{reactions~(\ref{reaction_hp})}, (\ref{reaction_hocp}), (\ref{reaction_hocp-iso}), and for photoionization of H$_2$($v''$$\geq$4). 
This represents our complete reference model of \mbox{d203-506}.
The dotted curves
correspond to models that neglect H$_2$($v''$$\geq$4) photoionization, which we find to play a minor role at the high gas densities considered. However, this process plays a more dominant role at lower densities due to the more nonthermal relative populations of \mbox{FUV-pumped} \mbox{H$_2$($v'' \geq 4$)} levels (see models in \mbox{Appendix~\ref{appendix-Orion Bar}}).
Dashed curves correspond to models neglecting H$_2$ photoionization and using the thermal rate constants computed from the state-to-state rates. Again, this approximation is sufficiently accurate for 
\mbox{$n_{\rm H}$\,=\,10$^{7}$\,cm$^{-3}$}, but becomes much less reliable in lower-density gas (see \mbox{Fig.~\ref{fig:chemical_rates_H3p_Bar_1}}).

The right panels of Fig.~\ref{fig:chemical_rates_H3p_d203-506} show the formation rate of H$_2^+$ from different reactions (upper box), as well as the percentage contribution of different reactions to the total H$_3^+$ formation rate (lower box).
In the predominantly atomic PDR (zone~$i$), H$_3^+$ formation is dominated by the reaction \mbox{H$_2$ + H$_2^+$}, but the primary formation pathway for H$_2^+$ is the reaction \mbox{H$_2$ + H$^+$} -- with a significant supply of H$^+$ from (photo)chemical reactions -- along with a minor contribution from the photoionization of \mbox{H$_2^*$ ($v \geq 4$)}.
The resulting H$_3^+$ abundance in zone~$i$ is not very high because destruction reactions with H atoms and dissociative recombinations  with electrons are both very efficient.
 In terms of local abundances, this thin zone is characterized by \mbox{$x$(H$^+$)\,$\simeq$\,a few 10$^{-5}$\,$\gg$\,$x$(H$_2^+$)\,$\gtrsim$\,$x$(H$_3^+$)\,$\simeq$\,10$^{-12}$}.
This further implies that, even in the absence of an ionization front,
\mbox{narrow} hydrogen recombination lines could still trace the neutral PDR \citep[see observations by][]{Boyden25}.

Near the  H/H$_2$ dissociation front (disk zone~$ii$), H$_3^+$ reaches its peak abundance
($\sim$3$\times$10$^{-8}$) and column density ($\simeq$\,8.5\,$\times$\,10$^{12}$\,cm$^{-2}$). The $x$(H$_3^+$) profile closely follows that of HOC$^+$ and, consequently, that of \mbox{water vapor}, which locally reaches very high abundances, \mbox{$x$(H$_2$O)\,$\sim$\,10$^{-5}$}, due to the high gas temperatures and the efficiency of \mbox{reactions~(\ref{reaction_oh}) and (\ref{reaction_h2o})} in converting atomic oxygen into H$_2$O.
 At the $x$(H$_3^+$) peak,  H$_3^+$ formation is driven almost exclusively by \mbox{H$_2$ + HOC$^+$} reactions. 
In our model, the majority of the H$_3^+$ column density originates from this zone. Its formation is largely independent of cosmic-ray ionization and instead depends on the abundance of HOC$^+$. 
 Indeed, the observed distribution of 
\mbox{HCO$^+$ $J$\,=\,4--3} emission in \mbox{d203-506} -- matching that of vibrationally excited H$_2^*$ \citep{Berne24}, highly rotationally excited OH \citep[i.e., the product of water vapor photodissociation;][]{Zannese24}, and fluorescent C\,\textsc{i} emission \citep{Goico24} -- implies that HOC$^+$, and thus H$_3^+$, originate from this PDR component.

In the more FUV-shielded and colder molecular gas, where \mbox{$f$(H$_2$)\,$\rightarrow$\,1} (disk zone~$iii$), most gas-phase carbon becomes locked in CO. As the gas temperature and FUV flux decrease, conditions begin to favor the formation of abundant water ice mantles. Consequently, the gaseous C/O abundance ratio increases. 
Owing to strong external irradiation -- leading to ice-mantle photodesorption and dust grain 
heating -- the vertical position of the water snow line is shifted deeper inward \mbox{\citep[e.g.,][]{Goico24}}  compared to that in isolated disks \mbox{\citep[e.g.,][]{Oberg11,Oberg16}}.
In terms of the ionization fraction, $x$(H$^+$) and $x$(e$^-$) decrease drastically in this zone. Here, cosmic rays, X-rays, and perhaps the decay of short-lived radionuclides become the only significant sources of ionization \citep{Cleeves13,Cleeves15}. H$_3^+$ forms through reactions between H$_2$ and H$_2^+$. However, the local H$_3^+$ abundances are lower -- approximately $10^{-11}$ for our adopted value of $\zeta$ -- than in the \mbox{FUV-irradiated} layers, due to less efficient formation and enhanced destruction by
proton transfer reactions with CO, N$_2$, and other species. 
The column density of cold H$_3^+$ in the midplane of a protoplanetary disk remains uncertain, but its IR ro-vibrational emission is expected to be faint.
In any case, JWST observations of \mbox{d203-506} are not sensitive to the cold, deep midplane layers or to the inner regions of the disk near the star.

\begin{table*}[!h] 
\begin{center}
\caption{Spectroscopic parameters  of the IR H$_3^+$ lines reported in d203-506 and  \texttt{GROSBETA} model predictions (see \mbox{Appendix~\ref{appendix-grosbeta}}).}  \label{Table_intensities}  
\normalsize
\begin{tabular}{c c c c c c c c @{\vrule height 10pt depth 5pt width 0pt}}    
\hline\hline       

Line$^a$                                        &  Band$^a$                           &  $\lambda$  & $A_{\rm ul}$ & $E_{\rm u}$/$k_{\rm B}^{\,b}$  & $T_{\rm ex}^{model}$ & $F_{\rm line}^{\,model,\,c}$ & $F_{\rm line}^{\,obs,\,d}$ \\  
   $P\,|Q\,|R\,$($J''$,\,$G''$)$^{u,l}$                                         &   $\nu'_{1}\,{\nu'_2}^{|l'|}$\,$\rightarrow$\,$\nu''_{1}\,{\nu''_2}^{|l''|}$                           &  ($\mu$m)   & (s$^{-1}$)    &                (K)         &     (K)      &  (W\,m$^{-2}$)         &  (W\,m$^{-2}$) \\ \hline
$Q(5,0)$ &  01$^1$\,$\rightarrow$\,00$^0$ &   4.045     &  1.13E$+$02  &  5293.3                  &   274.6      &  6.5E$-$22              &   1.7E$-$21      \\ 
$Q(5,1)^l$ &  01$^1$\,$\rightarrow$\,00$^0$ &   4.045     &  1.09E$+$02  &  5263.8                  &   273.4      &  3.1E$-$22              &   blended      \\ 
$Q(5,3)^l$ &  01$^1$\,$\rightarrow$\,00$^0$ &   4.044     &  7.18E$+$01  &  5020.2                  &   261.2      &  3.5E$-$22              &   blended      \\ 
$Q(5,2)^l$ &  01$^1$\,$\rightarrow$\,00$^0$ &   4.043     &  9.50E$+$01  &  5174.2                  &   269.0      &  2.6E$-$22              &   blended      \\ 
$Q(1,0)$ &  01$^1$\,$\rightarrow$\,00$^0$ &   3.953     &  1.28E$+$02	&  3672.6                  &   257.9      &  4.2E$-$22              &   2.3E$-$21    \\                                                                         
$R(1,0)$ &  01$^1$\,$\rightarrow$\,00$^0$ &   3.669     &  9.79E$+$01	&  3954.8                  &   272.6      &  4.4E$-$22              &   2.4E$-$22    \\
$R(7,6)^u$ &  01$^1$\,$\rightarrow$\,00$^0$ &   3.052     &  1.01E$+$02	&  6904.2                  &   329.5      &  2.8E$-$22              &   3.8E$-$22 \\ 
$R(7,0)$ &  01$^1$\,$\rightarrow$\,00$^0$ &   3.022     &  1.74E$+$02	&  8006.7                  &   427.4      &  2.8E$-$22              &   7.2E$-$22 \\ 
  $J',K'$\,$\rightarrow$\, $J'',K''$     &          &       &  	&                    &         &                &    \\\hline  
(5,0)\,$\rightarrow$\,(4,3)              &  00$^0$\,$\rightarrow$\,00$^0$  &  16.325     &  3.01E$-$03  & 1736.8                   &   750.8      &  1.8E$-$21              &  $<$\,6.2\,E$-$21 (3$\sigma$) \\\hline
\end{tabular} 
\end{center} 
\normalsize
\vspace{-0.5cm}
\tablefoot{$^a$Following \citet{Linsday2001}. $^b$Determined from the lowest accessible level of H$_3^+$, $J,K = (1,1)$.
$^c$Computed using an aperture of 0.1$''$ (a solid angle of \mbox{2.35$\times$10$^{-13}$\,sr}), the same aperture  used to extract the observational data}. $^d$From \citet{Ilane25b}.                
\end{table*}      

\subsection{H$_3^+$ excitation and infrared  emission in d203-506}
\label{sec:excitation}

\citet{Ilane25b} estimated an H$_3^+$ column density of \mbox{$\simeq 10^{13}$\,cm$^{-2}$}, based on an local thermodynamic equilibrium  (LTE)  fit to their likely detection of $\nu_2$ ro-vibrational lines in \mbox{d203-506}. This corresponds to an abundance $x$(H$_3^+$) of a few $10^{-8}$, using $N_{\rm H}$ in the disk PDR derived from JWST observations of multiple rotational lines of  \mbox{H$_2$($v$\,=\,0)}  \mbox{\citep[e.g.,][]{Berne23,Berne24}}. These 
values are broadly consistent with our PDR model results, with most of the H$_3^+$
column density arising from PDR  zone~$ii$ (Figs.~\ref{fig:chemical_model_d203-506} and 
\ref{fig:chemical_rates_H3p_d203-506}). \mbox{However}, it raises questions about how the $\nu_2$ bending mode is actually excited, given the very large 
Einstein $A_{ij}$ coefficients of the observed ro-vibrational lines \citep[$\simeq$100\,s$^{-1}$; e.g.,][]{Mizus17,Bowesman23} and the correspondingly high critical densities 
\mbox{($n_{\rm cr}$\,$\approx$\,10$^{13}$\,cm$^{-3}$)} required for collisional exitation -- much higher than the typical  density in the disk PDR, 
\mbox{$n_{\rm H}$\,$\approx$\,$10^7$\,cm$^{-3}$}.

Due to the low densities (compared to $n_{\rm cr}$) and the short chemical lifetime of H$_3^+$ in the disk PDR -- on the order of a few hours and dominated by destruction reactions with electrons and H atoms -- H$_3^+$ vibrational levels are unlikely to be thermally populated at $T_{\rm gas}$ through inelastic collisions.
Thus, accurate modeling of the vibrational excitation  requires one to incorporate the H$_3^+$
chemical formation and destruction rates (\mbox{$F$ and $D$}, respectively; see \mbox{Appendix~\ref{appendix-grosbeta}}) into the statistical equilibrium equations governing the ro-vibrational level populations. In particular, when
 \mbox{$n_{\rm H} \ll n_{\rm cr}$} and radiative pumping is not relevant  \citep[see also][]{Zannese25}, the excess energy from an \mbox{exoergic} formation pathway can leave the nascent H$_3^+$ molecule in a 
 vibrationally excited state, thereby driving much of the observed IR \mbox{ro-vibrational} emission.

To test whether the predicted physical conditions and H$_3^+$ chemistry in the PDR component of \mbox{d203-506} can reproduce the reported H$_3^+$ line intensties, we used the \mbox{single-slab} escape-probability code \texttt{GROSBETA} \citep{Black98,Tabone21,Zannese24},
an enhanced version of \texttt{RADEX} \mbox{\citep{Tak07}}, to compute a non-LTE excitation model of H$_3^+$. 
This code solves the statistical equilibrium equations by accounting for local chemical formation and destruction rates, \mbox{ro-vibrational} collisional excitation, and spontaneous emission.
We assume that the \mbox{ro-vibrational} level populations of the nascent H$_3^+$ molecule follow a Boltzmann distribution at an effective formation temperature, $T_{\rm form}$.
The input parameters are approximately those of disk \mbox{zone~$ii$}: 
\mbox{$n$(H$_2$)\,=\,$n$(H)\,=\,5$\times$10$^6$\,cm$^{-3}$}, \mbox{$T_{\rm gas}$\,=\,1{,}000\,K},  and
\mbox{$N$(H$_3^+$)\,=\,10$^{13}$\,cm$^{-3}$} 
(see \mbox{Appendix~\ref{appendix-grosbeta}} for more details and for the implemented  inelastic collisional rates).
 
Based on our PDR model results, H$_3^+$ formation is assumed to be dominated by reaction~(\ref{reaction_hocp}) 
when H$_2$ is in its ground vibrational state ($v''$\,=\,0)\footnote{The contribution of H$_2$($v'' \geq 1$) levels to H$_3^+$ formation via \mbox{reaction~(\ref{reaction_hocp})}  is less than 10\%~at the H$_3^+$ column density peak (see Fig~\ref{fig:HOC+rates_PDR}).}. 
Thus, we adopt \mbox{$F$(H$_3^+$)\,$\simeq$\,$k_9$($v''$=0)\,$\cdot$\,$n$(H$_2$;\,$v''$=0)\,$\cdot$\,$n$(HOC$^+$)
$\simeq$\,10$^{-5}$\,cm$^{-3}$\,s$^{-1}$}  from the reference PDR model (see Fig~\ref{fig:HOC+rates_PDR}), and we vary $T_{\rm form}$ as an input parameter.
Adopting $T_{\rm form}$\,=\,3{,}000\,K, the resulting non-LTE excitation model 
(see \mbox{Table~\ref{Table_intensities}}) matches reasonably well the intensities of the unblended H$_3^+$ lines
reported by \citet{Ilane25b}. Furthermore, a population diagram fit to the synthetic lines from the most populous levels gives \mbox{$T_{\rm rot}$(H$_3^+$) $\simeq$ 1{,}100\,K}, which is on the same order as the LTE excitation temperature range estimated by \citet{Ilane25b}. Therefore, our chemical and excitation models support the detection of \ch{H3+} emission in the PDR component of d203-506 and suggest that \ch{H3+} may be detectable in other strongly irradiated disks.

Since \mbox{reaction~(\ref{reaction_hocp})} is endoergic by 1{,}500\,K 
when H$_2$ is in the ground vibrational and rotational state (\mbox{$v''$\,$=$\,0} and \mbox{$J''$\,$=$\,0}), and \mbox{$T_{\rm form} \approx E_{\rm H_2}(v''=0, J'')/k - 1{,}500$\,K} (where $E_{\rm H_2}(v''=0, J'')/k$ is the energy of the rotational level $J''$ in Kelvin),  
 this model implies that H$_3^+$ formation in the disk PDR primarily proceeds via exoergic reactions of rotationally excited \mbox{H$_2$($v''$\,=\,0,\,$J''$\,$\geq$7)} with HOC$^+$. 
 Given the relatively high densities and temperatures in the PDR component of \mbox{d203-506}, these rotationally excited  \mbox{H$_2$($v''$\,=\,0)} levels are significantly populated\blue{$^{2}$}. Indeed, their associated IR emission lines are readily detected, corresponding to a rotational temperature of $\sim$950\,K
\citep[e.g.,][]{Berne23,Zannese24}, which is very close to the predicted gas temperature in 
\mbox{zone~$ii$} of the disk PDR
(upper panel of Fig~\ref{fig:chemical_model_d203-506}). Consistent with this scenario, our PDR model predicts substantial populations of \mbox{H$_2$($v''$\,=\,0,\,$J''$\,$\geq$\,7)} 
in this zone (see the upper left panel of \mbox{Fig.~\ref{fig:chemical_rates_H3p_d203-506}}), where the abundance of H$_3^+$  peaks.

\subsection{H$_3^+$ column density  as a function of external $G_0$, $\zeta$,
and dust grain properties}

\begin{figure}[t]
\centering   
\includegraphics[scale=0.47, angle=0]{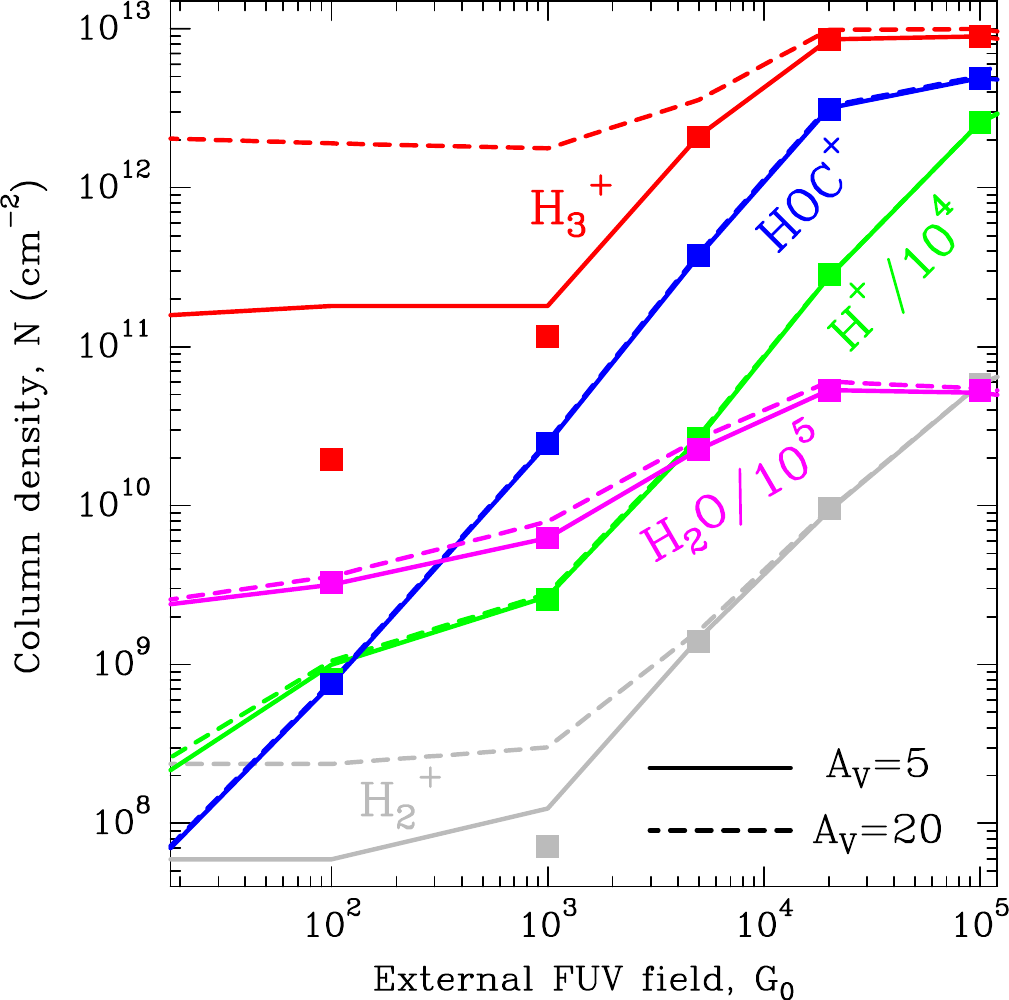}
\caption{Column densities of  H$^+$, H$_2^+$, H$_3^+$, H$_2$O, and HOC$^+$ as a function of external FUV field for models
with \mbox{$\zeta$\,=\,10$^{-16}$\,s$^{-1}$}. Continuous curves refer to column densities from $A_V$\,=\,0 to 5\,mag (tracing the disk PDR), whereas dashed curves  integrate up to 20\,mag
(tracing colder gas).  The filled squares refer to $A_V$\,=\,0 to 5\,mag models with 
\mbox{$\zeta$\,=\,10$^{-17}$\,s$^{-1}$}.} 
\label{fig:Grid}
\end{figure}

To isolate the role of increasing $G_0$ and varying $\zeta$ cosmic-ray ionization rates, \mbox{Fig.~\ref{fig:Grid}} shows the results of a grid of constant-density models
(\mbox{$n_{\rm H} = 10^7$\,cm$^{-3}$}) for different
strengths of the external FUV field and for a constant \mbox{$\zeta = 10^{-16}$\,s$^{-1}$}. This plot shows increasing column densities of H$^+$, H$_2^+$, H$_3^+$, H$_2$O, and HOC$^+$ as a function of  $G_0$.  Among these species,  \mbox{$N$(HOC$^+$)} is a particularly good tracer of $G_0$.
The continuous curves represent column densities integrated up to \mbox{$A_V = 5$\,mag} 
(\mbox{i.e., FUV-irradiated} gas in the disk PDR), whereas the dashed lines show column densities integrated up to 
\mbox{$A_V = 20$\,mag} (i.e., including \mbox{FUV-shielded} gas representative of the cold disk midplane). When the two curves are close, it indicates that most of the column density of a given species arises from FUV-irradiated gas, near the disk surface, and not from the cold midplane.
For $G_0 < 10^3$, the column densities of these species increase slowly with $G_0$. 
In this lower-irradiation regime, most of the H$_3^+$ and H$_2^+$ column density originates from colder gas, partially FUV-shielded, whereas H$^+$, HOC$^+$, and H$_2$O 
continue to primarily trace the warmer outer disk surface and wind.
We recall that our models do not apply to the much denser and hot inner disk regions close to the host star ($r <$ a few astronomical units), where water vapor is highly abundant 
\citep[with \mbox{$N$(H$_2$O)\,$\gtrsim$\,10$^{18}$\,cm$^{-2}$}; e.g.,][]{Carr08,Pontoppidan10,Riviere12,Du14,vanDishoeck21,Bosman22,Banzatti23,vanDishoeck23,Smith25}.

The curves in Fig.~\ref{fig:Grid} represent models with the same value of $\zeta$. Nevertheless, $N(\mathrm{H}^+)$, $N(\mathrm{H}_2^+)$, and $N(\mathrm{H}_3^+)$ increase significantly with $G_0$, particularly in the high-irradiation regime ($G_0 > 10^3$).
This trend is a clear signature of FUV-driven chemistry and \mbox{ionization}.
 In the disk's PDR component, the column density of these species trace the strength of the FUV  field, which triggers an active high-temperature photochemistry.
To assess the role of $\zeta$ in this PDR chemistry, the filled squares in Fig.~\ref{fig:Grid} represent column densities -- integrated up to $A_V = 5$\,mag -- from models with 
\mbox{$\zeta = 10^{-17}$\,s$^{-1}$} (a tenfold decrease in the ionization rate). 
In the PDR component of a strongly irradiated disk ($G_0 \gtrsim 10^3$), models with different values of $\zeta$ yield very similar results. Therefore, the dependence on $\zeta$ is very weak.

The predicted water vapor column densities in the disk PDR
($\sim$10$^{14}$ to several 10$^{15}$\,cm$^{-2}$) are not high enough 
 to reduce the FUV opacity through water self--shielding \citep{Bethell09,Bosman22},
and are also insufficient to produce detectable IR H$_2$O rovibrational emission due to very subthermal collisional excitation and ongoing photodissociation \citep[][]{Zannese24}.
Nonetheless, any dynamical process occurring on timescales comparable to the \mbox{chemistry -- such} as the advection of water-ice--coated grains into the PDR surface by the photoevaporative flow \citep[e.g.,][]{Maillard21,Coleman25}, followed by ice-mantle photo- and thermal-desorption \citep[e.g.,][]{Walsh13,Portilla-Revelo25}\mbox{ -- could} enhance the abundance of water vapor at the outer disk surface, and consequently increase the abundances of HOC$^+$ and H$_3^+$ beyond our predictions.
Alternatively, efficient grain growth, settling to the midplane, and radial drift may progressively remove these icy grains from the \mbox{upper} disk, reducing the gaseous oxygen 
 abundance in the PDR  \mbox{\citep[e.g.,][]{Oberg16,Coleman25}} and, indirectly, the abundance of H$_3^+$. 

The adopted grain-size distribution also impacts the FUV opacity, and thus the penetration of external FUV radiation. 
Our choice of the size distribution exponent and the $a_{\rm min}$ and $a_{\rm max}$ radii 
(\mbox{Sect.~\ref{sec:PDR_reference}}) 
is consistent with the modest grain growth expected in the upper layers of a young irradiated disk 
\mbox{\citep[e.g.,][]{Stoerzer99}}. 
However, over time, the smallest dust grains are expected to be entrained in the photoevaporative wind \citep[e.g.,][]{Facchini16}. 
Both grain growth and dust entrainment reduce $\sigma_{\rm ext}^{\rm d}$ (and increase the dust albedo and anisotropy of the scattered FUV radiation), 
thereby lowering the FUV opacity and enhancing the penetration of FUV photons  
\mbox{\citep[][]{Goico07}}. A lower $\sigma_{\rm ext}^{\rm d}$ value results in a more extended C$^+$ layer and an increased HOC$^+$ column density. 
However, the higher electron density at the H$_3^+$ abundance peak enhances its destruction, thereby slightly reducing the H$_3^+$ column density. A model with
more substantial grain growth, \mbox{$\sigma_{\rm ext}^{\rm d}$\,$=$\,$3.5 \times 10^{-22}$\,cm$^2$\,H$^{-1}$} at 1{,}000\,\AA\,, results in a factor $\sim$1.3 lower $N$(H$_3^+$)
than the reference model.
Detections of H$_3^+$ in other  irradiated disks will help to constrain these different scenarios.

\subsection{Additional plausible pathways for H$_3^+$ formation}

Several studies have explored alternative H$_3^+$ formation mechanisms in other astrophysical environments. Most of these mechanisms involve reactions with electronically excited states of atomic hydrogen and molecular hydrogen; for instance,
\begin{equation}
{\rm H_2^*\,(Rydberg)}\,+\, {\rm H_2}\,\rightarrow\,{\rm H_3^+}\,+\,{\rm H}\,+\,{\rm e^-},
\label{reaction_chemiionization}
\end{equation}
which is a chemiionization process, as well as
\begin{equation}
{\rm H^*\,(\textit{n}\,=\,2, ...)}\,+\, {\rm H_2}\,\rightarrow\,{\rm H_3^+}\,+\,{\rm e^-},
\label{reaction_associative_ion}
\end{equation}
which is an associative ionization \citep[e.g.,][]{Dehmer95}. In addition, associative 
ionization  involving ground-state and excited states of atomic hydrogen 
may contribute. That is,
\begin{equation}
{\rm H\,(1s)}\,+\, {\rm H(\textit{n}\,>\,2)}\,\rightarrow\,{\rm H_2^+}\,+\,{\rm e^-},
\label{reaction_associative_ion2}
\end{equation}
which is endoergic by $\sim$12,770\,K. The importance of this process was emphasized by \citet{Rawlings93}, and there is both experimental and theoretical support \citep[][]{Urbain91,Urbain92,Hornquist23}.
The rate coefficient is small for $n = 2$ at temperatures of interest, but for $n > 2$, the process is relatively fast (the reaction becomes exoergic). Its contribution to the production of H$_2^+$ and H$_3^+$ will be limited by the populations of high-$n$ states of atomic hydrogen.
In the context of protoplanetary disks externally irradiated by both EUV and FUV radiation, recombination near the ionization front might contribute to H($n > 2$). However, we suspect that the resonant trapping of Lyman line photons in a very thin layer of the PDR might be a bigger effect in populating excited states. While these associative ionization processes likely produce lower H$_3^+$ column densities compared to the photochemistry and  reactions discussed in the previous sections, they may play a larger role in environments  such as centers of active or star-forming galaxies. In particular, X-ray- or cosmic-ray-dominated galaxies would naturally have a thicker zone where H, H($n > 2$), and H$^+$ coexist than star-powered (PDR) galaxies.

All in all, \mbox{UV- and X-ray-driven}  H$_3^+$ formation may be relevant also in the inner disk regions (near the host star) as well as in other environments such as exoplanet atmospheres -- where H$_3^+$ acts as an important coolant \citep[e.g.,][]{Koskinen07,Khodachenko15} -- and the early Universe \citep[e.g.,][]{Lepp02,Glover03,Coppola11,Coppola13}. 
Our state-dependent rate constants (Table~\ref{table:rates}) may be valuable for modeling these environments and support the prospect of new H$_3^+$ detections with JWST \citep[e.g.,][]{Richey-Yowell25}.

 \section{Summary and conclusions}

We have presented a detailed photochemical model of the PDR component of a protoplanetary 
disk -- comprising the outer disk surface and photoevaporative wind -- exposed to strong external FUV radiation. We have revisited key \mbox{FUV-driven} gas-phase reactions leading to H$_3^+$ formation, including new state-to-state dynamical calculations of the vibrational-state $v$-dependent rate constants for reactions of H$_2$($v$) with HOC$^+$ and H$^+$. In addition, we have modeled the   photoionization of vibrationally excited 
\mbox{H$_2$($v \geq 4$)} by FUV, a process new to disk and PDR models.

These reactions dominate the formation of H$_3^+$ in the disk PDR, 
largely independent of the cosmic-ray ionization rate. We conclude that the IR line emission reported by \citet{Ilane25b} is consistent with the presence of H$_3^+$ in the PDR component of \mbox{d203-506}. However, our results indicate that H$_3^+$ is not a reliable tracer of $\zeta$ in this component, but instead primarily traces the strength 
 of the external FUV field ($G_0$) and gas-phase photochemical processing.
Under the conditions of this disk PDR  (\mbox{$G_0 \simeq 2 \times 10^4$} and 
\mbox{$n_{\rm H} = 10^7$\,cm$^{-3}$}), we predict a peak H$_3^+$ abundance of $\gtrsim 10^{-8}$ in the irradiated disk surface -- corresponding to a column density of \mbox{$N$(H$_3^+$) $\lesssim 10^{13}$\,cm$^{-2}$} -- where HOC$^+$ is also abundant and plays a dominant role in driving the formation of H$_3^+$.
This reactive molecular ion  is ultimately linked to the enhanced abundances of C$^+$ \mbox{(i.e., due to  FUV radiation)} and H$_2$O \citep[see also][]{Portilla-Revelo25}.

Given the relatively low densities  (compared to  
the critical densities of the IR H$_3^+$ \mbox{$\nu_2=1 \rightarrow 0$}  lines) and the short chemical lifetime of H$_3^+$ in the disk PDR, H$_3^+$  rovibrational levels are unlikely to be thermalized by inelastic collisions with H$_2$, H, or e$^-$.
A coupled non-LTE excitation and chemical formation model, with the typical physical conditions of the  disk PDR  ($n_{\rm H}$\,$\simeq$\,10$^7$\,cm$^{-3}$ and $T_{\rm gas}$\,$\simeq$\,1{,}000\,K),
reproduces the IR H$_3^+$  line intensities
reported by \citet{Ilane25b},  provided that formation pumping following  
\mbox{exoergic} reactions 
drives most of the H$_3^+$ vibrational excitation. 
The model supports a scenario in which reactions between \mbox{rotationally} \mbox{excited} \mbox{H$_2$ ($v''$\,=\,0,\,$J''$\,$\geq$\,7)} and HOC$^+$ account for the reported H$_3^+$ line intensities in \mbox{d203-506}.

Our models show that the abundances of key molecular species (notably molecular radicals and reactive ions) are significantly enhanced in the PDR component of externally irradiated disks.
However, their exact abundance depends both on the exact $G_0$/$n_{\rm H}$ value and on grain and PAH properties, which are not well constrained in many disks. 
Furthermore, the observable IR ro-vibrational spectrum ultimately depends on the underlying excitation mechanism, which is often closely tied to the species' chemical formation and destruction. Upcoming spectroscopic surveys with JWST and ALMA, targeting 
tens of protoplanetary disks in clustered environments 
\citep[e.g.,][]{Planet25}, will reveal this chemistry
in larger disk samples. The presence of IR \ch{H3+} emission  will be another signature of FUV-driven chemistry and photochemical processing. It remains to be seen whether face-on disks also exhibit \ch{H3+} features from their inner regions near the host star.

\begin{acknowledgements}  
We thank the referee for the careful reading of our manuscript and the constructive report. We also thank the PDRs4All team for stimulating discussions over the past few  years.
  JRG and OR thank the Spanish MCINN for funding support under grants
\mbox{PID2023-146667NB-I00} and \mbox{PID2021-122549NB-C21}.
We thank the PCMI  of CNRS/INSU with INC/INP, co-funded by CEA and CNES.

\end{acknowledgements}

%
%

\bibliographystyle{aa}
\bibliography{references}

\begin{appendix}\label{Sect:Appendix}

\section{Application to lower-density PDRs}
\label{appendix-Orion Bar}

\begin{figure}[h]
\centering   
\includegraphics[scale=0.40, angle=0]{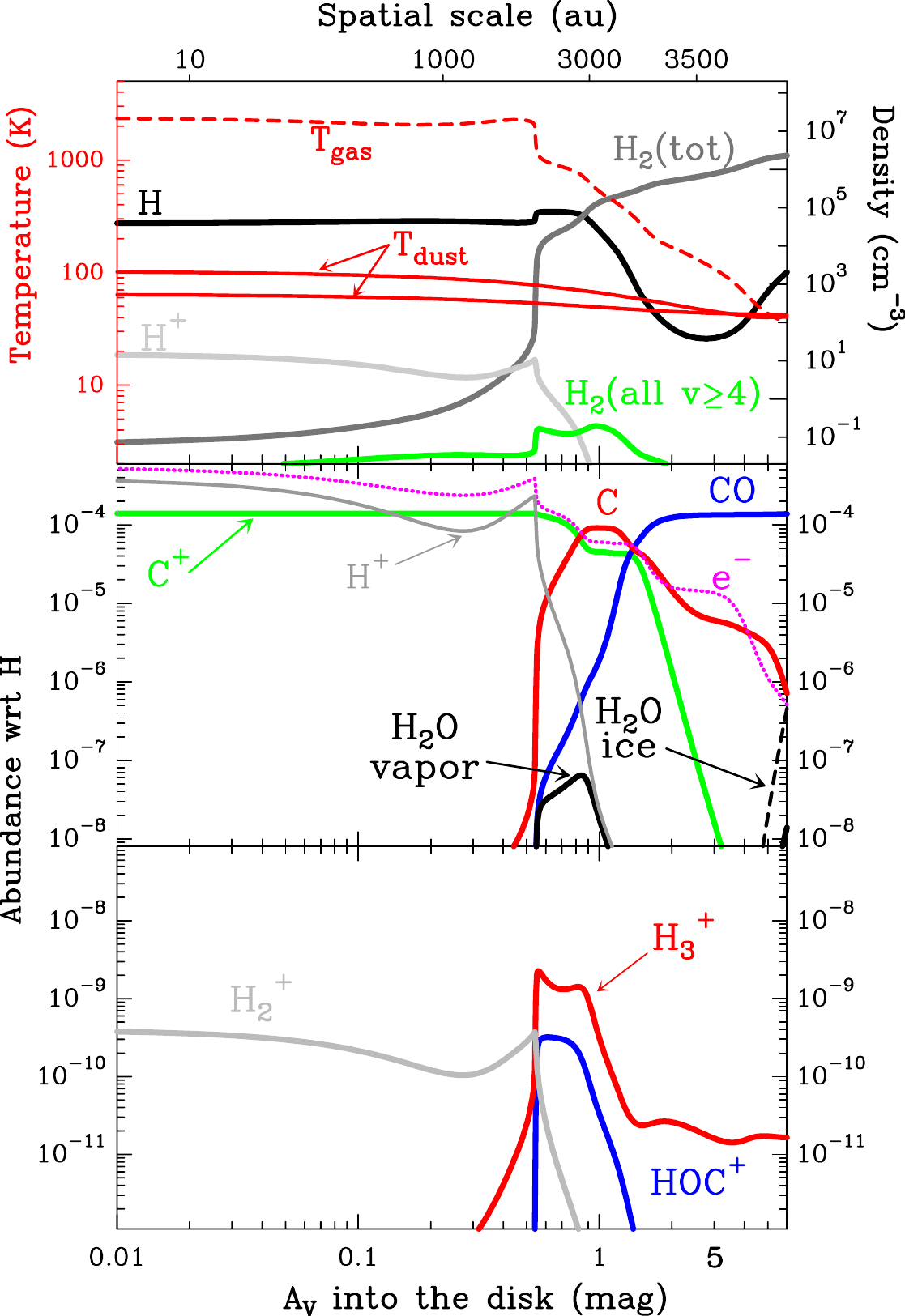}
\caption{Same as Fig.~\ref{fig:chemical_model_d203-506} but for the isobaric  model of the Orion Bar.}  
\label{fig:chemical_model_Orion Bar}
\end{figure}

\begin{figure}[t]
\centering   
\includegraphics[scale=0.41, angle=0]{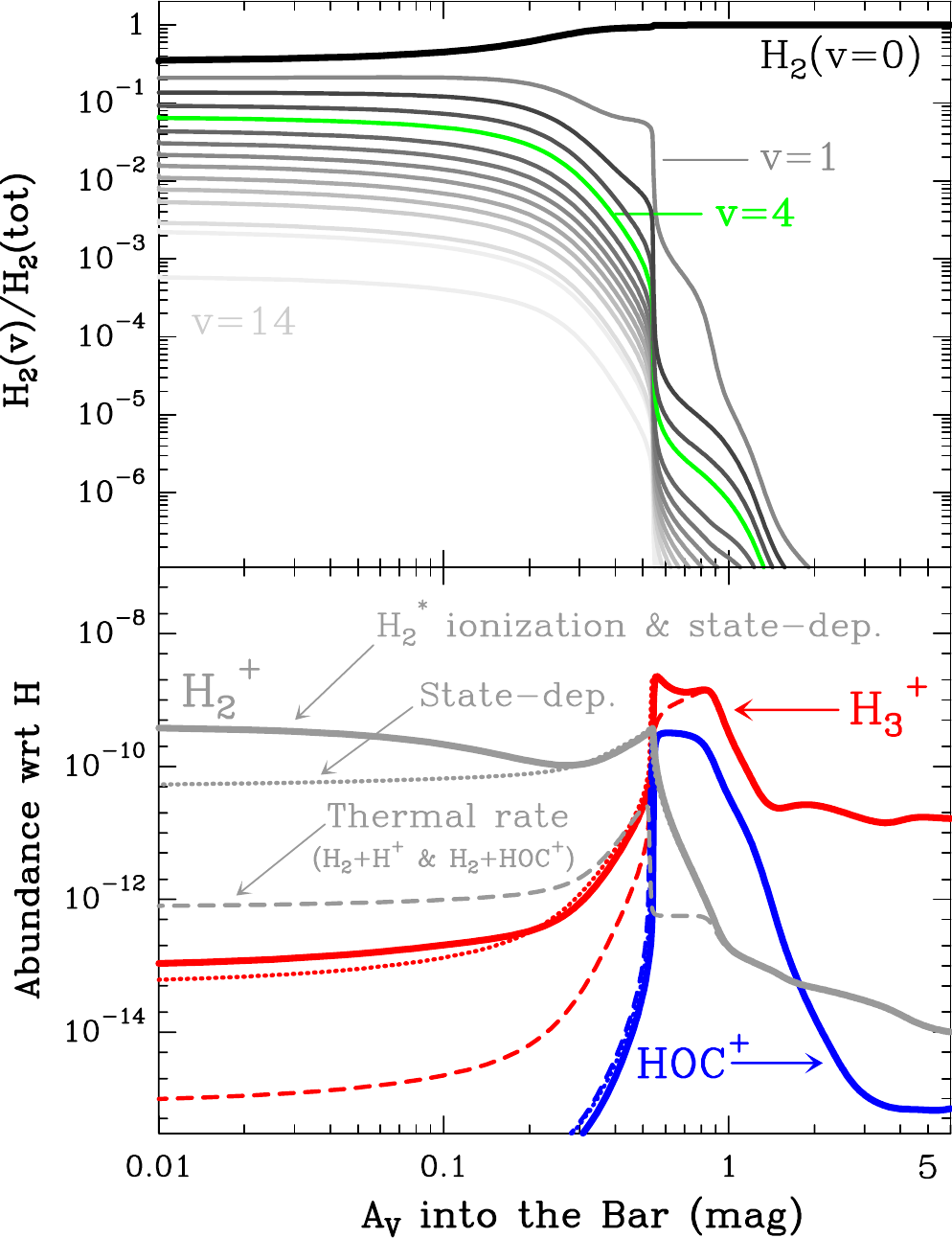}
\caption{Same as Fig.~\ref{fig:chemical_rates_H3p_d203-506} (left panels) but for the model of the Orion Bar.}  
\label{fig:chemical_rates_H3p_Bar_1}
\end{figure}

In this Appendix we model a lower-density PDR, resembling  either
interstellar PDRs such as the  Orion Bar or low-density photoevaporative winds.
In the Orion Bar, while $G_0$ is also a few 10$^4$, the gas density is lower, about 5$\times$10$^4$\,cm$^{-3}$ in the predominantly atomic PDR, and up to $\sim$10$^6$\,cm$^{-3}$ in the densest molecular gas condensations \mbox{\citep[e.g.,][]{Tielens93,Goico16}}.
Figure~\ref{fig:chemical_model_Orion Bar} shows the predicted structure of the PDR
(analogous to Fig.~\ref{fig:chemical_model_d203-506}) for an isobaric model (i.e.,~with a
density gradient)
with \mbox{$P_{\rm th}$\,=\,10$^{8}$\,K\,cm$^{-3}$}
and \mbox{$G_0$\,=2$\times$10$^4$} \citep[e.g.,][]{Joblin18}.
This lower gas density produces a more spatially extended atomic PDR, and a H/H$_2$ transition zone located about ten times deeper in $A_V$. 
In addition, the populations of the vibrationally excited H$_2$($v$) levels deviate further from thermal equilibrium populations, reaching higher abundances relative to H$_2$($v = 0$) than in the \mbox{d203-506} model due to the more prominent role of FUV pumping at higher $G_0/n_{\rm H}$ values (see the upper panel of Fig.~\ref{fig:chemical_rates_H3p_Bar_1}).
However, due to lower gas densities, lower shielding, and enhanced  photodissociation, the  water abundance  at the PDR surface is significantly less abundant, with $x$(H$_2$O)\,$\lesssim$\,10$^{-7}$.

\begin{figure}[t]
\centering   
\includegraphics[scale=0.41, angle=0]{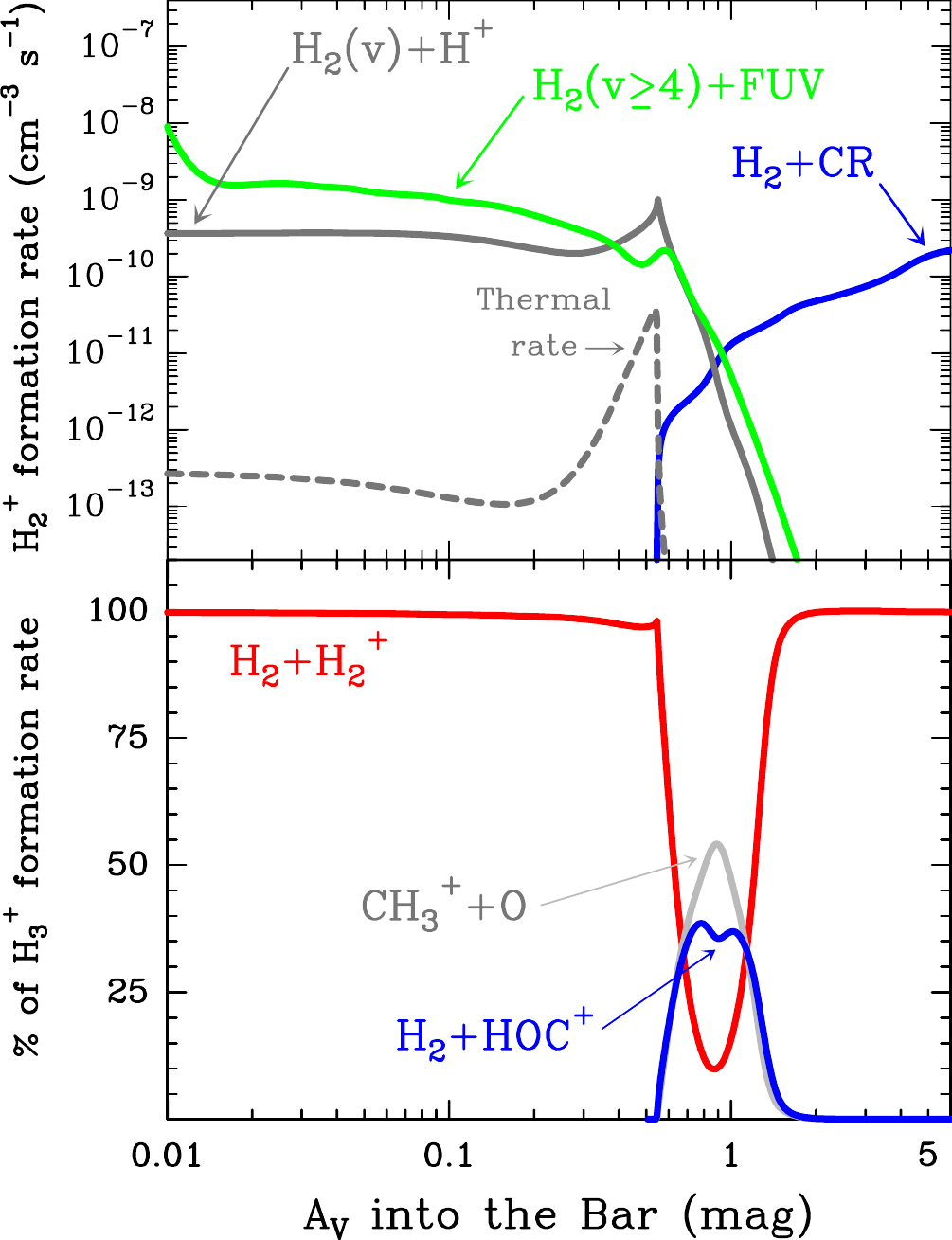}
\caption{Same as Fig.~\ref{fig:chemical_rates_H3p_d203-506} (right panels) but for the model of the  Bar.}  
\label{fig:chemical_rates_H3p_Bar_2}
\end{figure}

The more significant role of nonthermal H$_2$($v$) populations becomes evident when comparing the use of state-to-state versus thermal rates for reactions~(\ref{reaction_hp}), (\ref{reaction_hocp})
and (\ref{reaction_hocp-iso}). The latter approximation underestimates the 
H$_2^+$ and H$_3^+$ abundances by nearly two orders of magnitude in the atomic PDR zone.
Compared to the \mbox{d203-506} model, we find a less prominent role of the reaction 
\mbox{H$_2$($v$)\,+\,H$^+$}  in the surface layer of the PDR (partially due to lower gas temperatures), while H$_2$($v \geq 4$) photoionization becomes more significant. In addition, we predict \mbox{$x$(e$^-$)\,$>$\,$x$(C$^+$)} in this zone. 
In any case, most of the H$_3^+$ column density originates from deeper layers of the molecular PDR, where the peak $x$(H$_3^+$) abundance is a few $\times 10^{-9}$. 
This is more than an order of magnitude lower than the $x$(H$_3^+$) peak in the disk model, and is a result of the two orders of magnitude lower water vapor abundance, and thus lower HOC$^+$ abundance. 
Instead, the reaction
\mbox{CH$_3^+$ + O $\rightarrow$ H$_3^+$ + CO},
which is also related to the presence of C$^+$, CH$^+$, and H$_2^*$ \citep{Berne24,Zannese25,Goico25}, plays a more significant role 
(see \mbox{Fig.~\ref{fig:chemical_rates_H3p_Bar_2}}).
All in all, the total H$_3^+$ column density in this lower-density PDR model is $\sim$10$^{12}$\,cm$^{-3}$. Still, about 70\,\%\,of the $N$(H$_3^+$) column arises from FUV irradiated gas at $A_V < 1.5$\,mag. In this PDR zone, H$_3^+$
formation is not related to $\zeta$.

\clearpage

\section{Charge transfer H\,+\,H$_2^+$ $\leftrightarrow$ H$^+$\,+\,H$_2$}
\label{App:Charge_Tramsfer}

\begin{figure}[h!]
\begin{center} 
\includegraphics[width=0.86\linewidth]{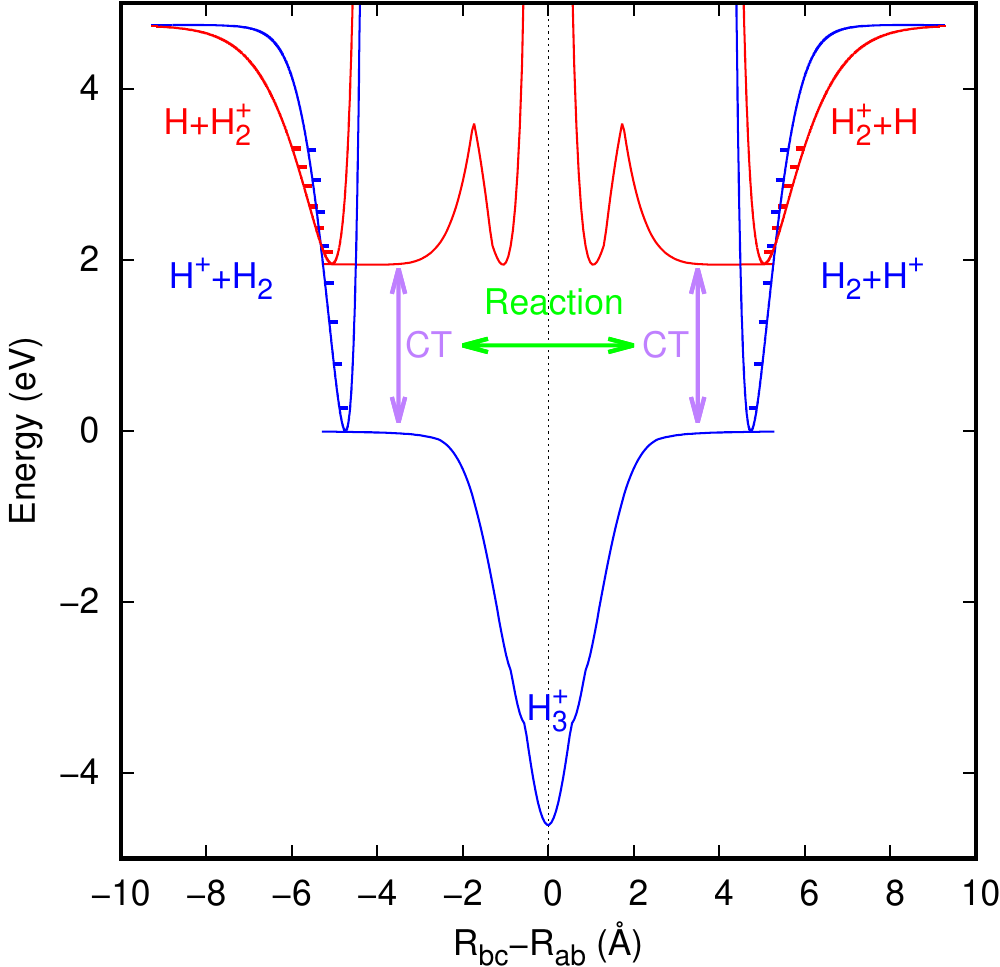}
\caption{{Minimum energy path of the two lower adiabatic  singlet electronic states of the H$_3^+$ system, correlating to H$_2$+H$^+$ (ground) and H$_2^+$ + H (excited). Vibrational levels of diatomic fragments  are indicated in blue (H$_2$) and red (H$_2^+$). The CT processes are described by electronic transitions between the two electronic state. Reaction only occurs in the ground electronic state.}}    
\label{fig:H3+meps}
\end{center}
\end{figure}

The CT reactions for the formation and destruction of H$_2^+$ take place in three 
electronic states, each one corresponding to a charge in one hydrogen atom, for an electron spin zero, the singlet states. The ground adiabatic electronic state presents a deep well (see Fig.~\ref{fig:H3+meps}), corresponding to the H$_3^+$ system, leading to \mbox{H$^+$\,+\,H$_2$} in the three possible rearrangement channels. The two adiabatic excited ones are repulsive and correlate to 
\mbox{H$_2^+$\,+\,H}, and present high barriers for the reaction. The direct and backward reactions can result in either simple CT, with no hydrogen exchange (non-reactive), or RCT, in which both charge transfer and hydrogen exchange occur. The latter involves products forming through a rearrangement channel different from the initial one, as illustrated in Fig.~\ref{fig:H3+meps} by the green horizontal line, which exists only in the ground adiabatic electronic state.

\subsection{Reaction H\,+\,H$_2^+(v=0,j=0)$\,$\rightarrow$\,H$^+$\,+\,H$_2$($v'$)}

\begin{figure}[h]
\begin{center}
  \includegraphics[width=0.99\linewidth]{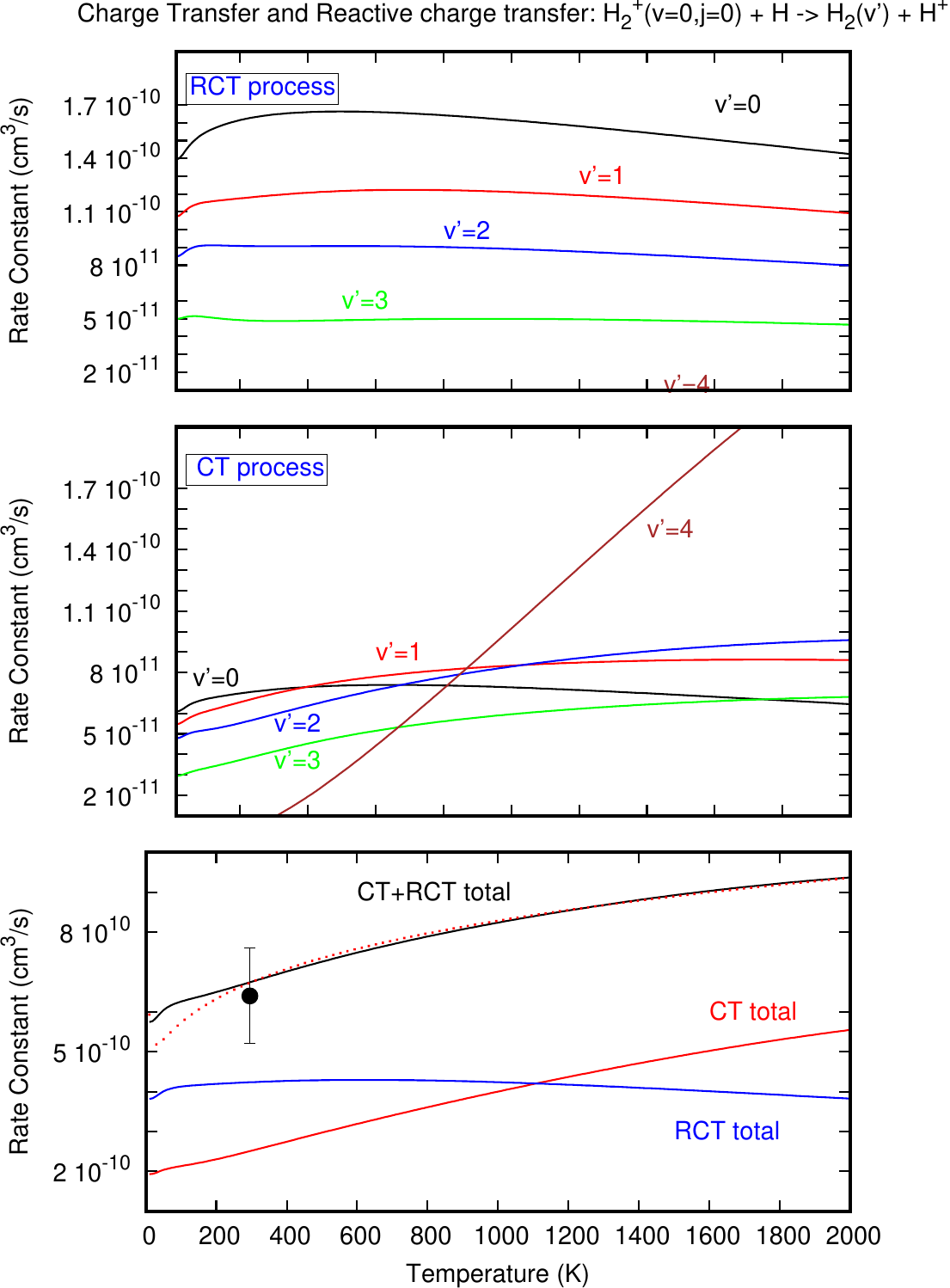}
  \caption{{State-dependent rate constants for the  reaction
  \mbox{H\,+\,H$_2^+(v=0,j=0)$$\rightarrow$H$^+$\,+\,H$_2$}. In the bottom panel, CT, RCT and the total sum are presented. In the middle and top panels, the pure CT and RCT process are shown, respectively, showing the final  state distribution of H$_2^+(v')$ vibrational states. 
The dotted curve represents the Arrhenius fit to the total rate constant.  
  The point in the lower panel corresponds to the experimental measurement by \cite{Karpas79}.}}    
\label{fig:H2formation}
\end{center}
\end{figure}

The cross section of the reaction \mbox{H\,+\,H$_2^+(v=0,j=0)$\,$\rightarrow$\,H$^+$\,+\,H$_2$} was obtained by \cite{Sanz-Sanz21} using a quantum wave packet method, based on the coupled electronic potential energy surfaces (PESs)  calculated by \cite{Aguado-etal:21}, where a detailed description of the reaction dynamics is provided. The state-dependent CT and RCT rate constants are obtained here by numerical integration over the collision energy with a Boltzmann energy distribution, and is presented in Fig.~\ref{fig:H2formation}. The total (CT+RCT) rate constant is good agreement with the experimental measurement at 300 K by \cite{Karpas79}. \cite{McCartney-etal:99} also studied this reaction, but at much higher energies of 30-100 keV, out of the scope of this study. 
More recently, \cite{Andrianarijaona-etal:09,Andrianarijaona-etal:19} studied the 
\mbox{H + D$_2^+$} isotopic
variant  in a broader energy range of 0.1-100 eV. Their results show a good agreement with
the  wave packet calculations performed using the same coupled 
 \citep[][]{Roncero-etal:22}.

Below 1{,}000\,K the RCT process dominates, involving a transition from the excited to the ground state, where the reaction takes place within the deep well due to the formation of H$_3^+(^1A')$ well.
Due to the formation of long-lived (H$_3^+$)$^*$ complexes, the reaction behaves nearly statistically \citep{Gonzalez-Lezana-etal:05,Gonzalez-Lezana-Honvault:17}. 

The vibrational distribution of H$_2(v')$ formed by the direct CT process shows some differences. $v'=$ 0,1 and 2 are very close to each other, while the $v'$ = 4 grows very fast once it is open at about 250 K. The reason is that the \mbox{H$_2(v'=4)$\,+\,H$^+$}  energy is very close to that of the  
\mbox{H$_2^+(v=0)$\,+\,H} reactants, and this small energy difference increases the efficiency of the non-adiabatic couplings between the two electronic states from very long distances, in the so-called resonant effect \citep{Sanz-Sanz21}.

Several approximate theoretical treatments have been applied to this reverse reaction \citep{Last-etal:97,Kamisaka-etal:02,Krstic02,Krstic-Janev:03,Errea-etal:05}, but the present calculations and results are considered the most accurate to date, as validated by the good agreement with the experimental value reported by \cite{Karpas79}.

\subsection{State-dependent rate constants for the endoergic reaction
H$_{2}\,(v'')$ + H$^+$\,$\rightarrow$\,H$_{2}^{+}$\,($v'$) + H}
\label{appendix-h2_hp}

The direct reaction H$^+$+H$_2$   has
been widely  studied at energies above the opening of the CT H + H$_2^+$ channel, at 1.82\,eV \citep{Markovic-Billing:95,Ichihara-etal:96,Last-etal:97,Chajia-Levine:98,Takayanagi-etal:00,Ushakov-etal:01,Ichiara00,Errea-etal:01,Kamisaka-etal:02,Krstic-Janev:03,Savin-etal:04,Kusabe-etal:04,Chu-Han:05,Urbain-etal:13,Sahoo-etal:14,Ghosh-etal:15}. Particularly relevant for this study are the studies by \cite{Ichiara00} and \cite{Krstic02}, in which this reaction was studied in a very large range of energies, up to a few eV, for different initial vibrational states of H$_2$, using approximate methods. \cite{Ichiara00}  used a quasi-classical (QCT) method including electronic transitions based in a \mbox{Landau-Zener} model, which is restricted to the crossing region between H$_2$+H$^+$ and H$_2^+$ + H (see Fig.~\ref{fig:H3+meps}). \cite{Krstic02} applied a quantum time independent close coupling method, but using an Infinite Order sudden (IOSA) approximation, which freezes the internal angle and decouples the helicities. Both approaches treat approximately the reactivity,
  making necessary a more accurate method for temperatures below 2{,}000\,K.

In this study, we use a quantum wave packet method that includes all degrees of freedom and can be considered ``exact'' for treating reactivity up to 2{,}000\,K. The calculations are similar to those performed for the reverse reaction \citep{Sanz-Sanz21}, as described in Appendix~\ref{App:Charge_Tramsfer}, and are carried out using the MADWAVE3 code \citep{Roncero-delMazo-Sevillano:25}. This code represents the Hamiltonian in reactant Jacobi coordinates within a body-fixed frame, employing grids for the internal coordinates (the Jacobi distances $r$ and $R$, associated with the H$_2$ internuclear vector and the vector joining the H$_2$ center of mass with H$^+$, respectively), and a basis for the electronic coordinates and total angular momentum ${\bf J}$, characterized by the quantum number $J$ and its projections $M$ and $\Omega$ on the space-fixed and body-fixed $z$-axes, respectively.

Due to the presence of the deep H$_3^+$ potential well, a large number of $\Omega$ projections is required to achieve convergence. In this work, we use $\Omega_{\mathrm{max}} = \min(J, 15)$. The flux into individual final states of the H$_2^+(v',j')$ products is evaluated using a reactant-to-product coordinate transformation method \citep{Gomez-Carrasco-Roncero:06}, applied separately for each $J$. To reduce computational cost, we explicitly calculate $J = 0$, 5, 10, 20, 30, 40, and 60. For intermediate $J$ values, the state-to-state reaction probabilities are interpolated using the $J$-shifting method \citep{Aslan-etal:12b}.

\begin{figure}[t]
\begin{center}
  \includegraphics[width=0.95\linewidth]{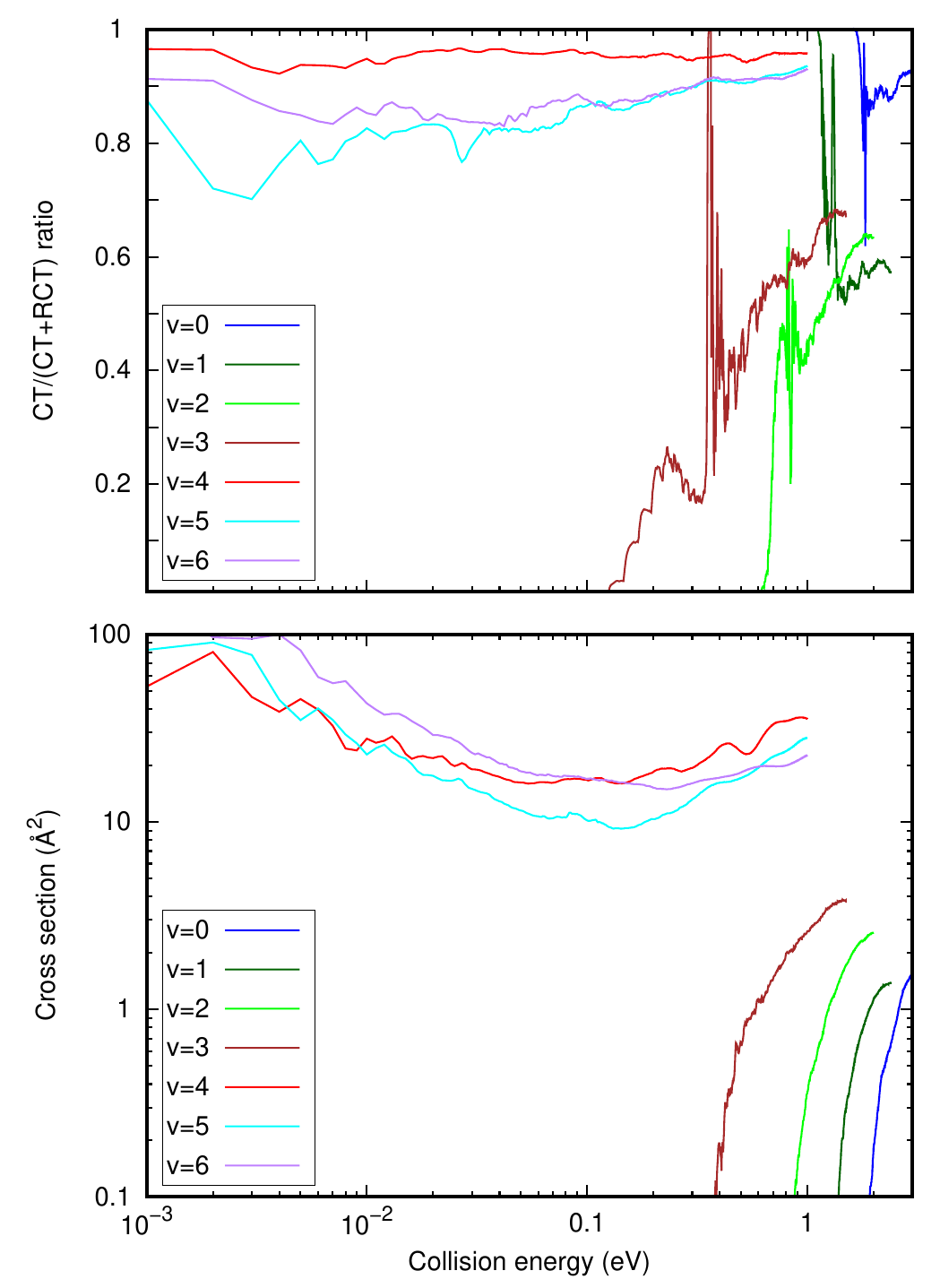}
  \caption{\textit{Upper panel:} Fraction of the CT cross section as a function of $v$.
  {\textit{Botom panel}: Total reactive cross sections (CT and RCT) for the reaction
  \mbox{H$^+$+H$_2(v)$ $\rightarrow$ H + H$_2^+$}  for several initial vibrational states $v$.}}    
\label{fig:H2-to-H2p-sigma}
\end{center}
\end{figure}

The total CT cross section (CT + RCT) for different initial H$_2(v,j=0)$ are shown in the bottom panel of Fig.~\ref{fig:H2-to-H2p-sigma}. The threshold for the CT is at 1.79 eV, so that the reaction is endothermic for $v < 4$, leading to negligible reactive cross sections below 0.4 eV for $v$=0-3 (see Fig.~\ref{fig:H3+meps}). However, for $v \ge 4$ the CT reaction becomes energetically accessible and the cross section becomes important, between 50 and 
200\,\AA$^2$ for $v$=4, 5 and 6. Qualitatively this result is similar to those reported previously by \cite{Ichiara00} in their \mbox{Fig.~1} and by \cite{Krstic02} in \mbox{Fig.~1}. However, 
we consider the magnitude of the exact quantum results obtained in this work is considerably more precise. For example, for $v=6$ below 2.5\,eV, the total cross section is always lower than 40 \AA$^2$ \citep{Ichiara00}, while here is between
100-200\,\AA$^2$. The large differences is attributed to classical character of the CT, made with a simple Landau-Zenner model. The IOSA results of \cite{Krstic02} for $v$=4 shows a threshold, which is absent here.

The CT/(CT+RCT) ratio shown in the top panel of Fig.~\ref{fig:H2-to-H2p-sigma} demonstrates that the dominant process is the simple CT without reaction. The RCT process acounts for a maximum of 30$\%$ for $v\ge$ 4, while is dominant for the lower $v<$ 4. This trend is also similar to the values reported by \cite{Ichiara00} in the low energy region. We find that for $v> 6$ the total CT cross section progressively diminish with increasing $v$. 
Moreover, it is difficult to extend the present quantum calculations calculations above $v=6$, because it is very close  to the three-body fragmentation channel, and we shall adopt here the behavior  by  \cite{Ichiara00}.

The CT rate constant are obtained by numerical integration of the cross section
with a Boltzmann distribution. \mbox{Figure~\ref{fig:New_rates_H2_Hp}}  shows the enhancement of the reaction rate as the vibrational excitation of the initial H$_2$ increases, becoming very similar for $v=4,5$ and 6, close to the Langevin value, with a rate constant of \mbox{$\simeq$\,(2-4)$\times$10$^{-10}$\,cm$^3$\,s$^{-1}$}.

\begin{figure}[t]
\begin{center}
\includegraphics[width=0.99\linewidth]{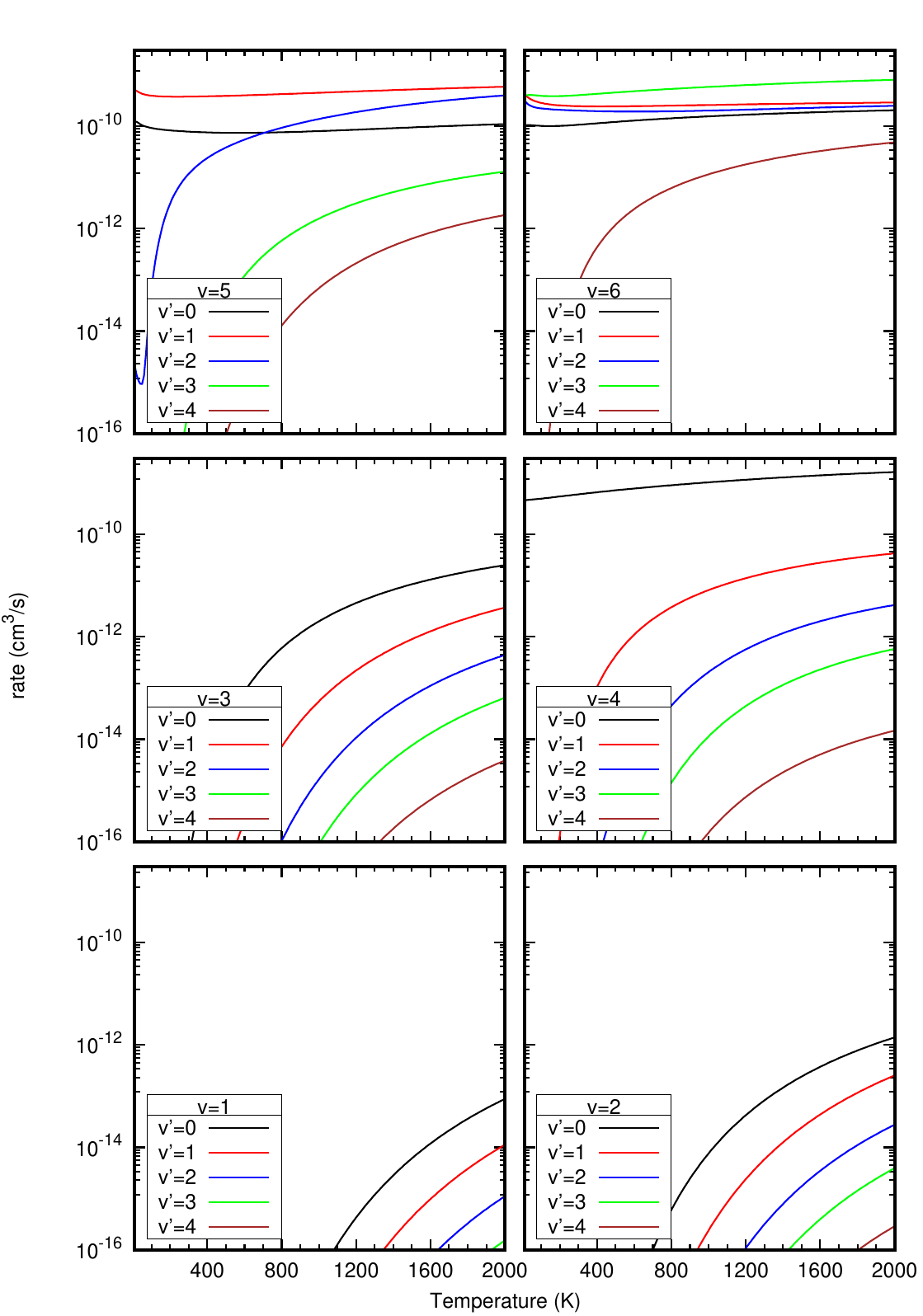}
\caption{State-to-state reactive rate constants (CT and RCT) for the 
reaction H$^+$+H$_2(v)$ $\rightarrow$ H + H$_2^+$($v'$) considering several initial $v$ and final $v'$ vibrational states. }  
\label{fig:finalvH2p}
\end{center}
\end{figure}

The final H$_2^+$($v'$) products populate progressively more excited vibrational states with increasing initial H$_2(v)$, as shown in Fig.~\ref{fig:finalvH2p}. For example, for an initial state of $v=6$, the dominant final state is $v'=3$. 
Moreover, the reaction \mbox{H$_2$\,+\,H$_2^+(v)$\,$\rightarrow$\,H$_3^+$\,+\,H} was recently studied for initial vibrational states $v=0$–6, and it was found that, up to collision energies of 1~eV, the reactive cross section is nearly independent of $v$, with values very close to the Langevin limit \citep{delMazo-Sevillano-etal:24a}. At higher collision energies, however, the reactive cross section increases with $v$.

\clearpage

\section{Photoionization of H$_{2}^{*}$\,($v''$\,$\geq$\,4)}
\label{appendix-photo-x-sections}

\begin{figure}[th]
\centering   
\includegraphics[scale=0.56, angle=0]{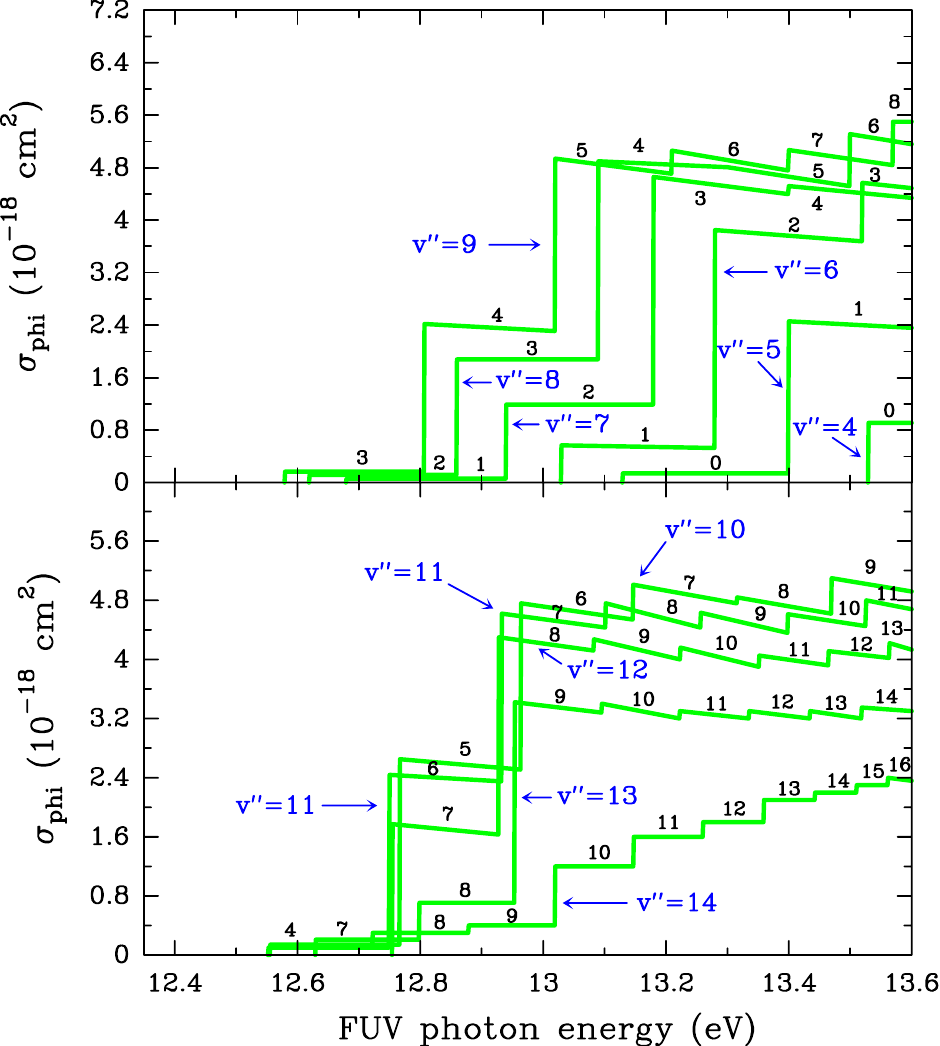}
\caption{Cross sections for FUV photoionization of H$_2^*$, from vibrationally excited levels $v''$ to H$_{2}^{+}$ vibrationally excited levels, $v'$,  as a function of FUV
photon energy \citep[from][]{Ford75}. Upper panel: \mbox{H$_2^*$($v''$\,=\,4 to 9)}. Lower panel:  \mbox{H$_2$($v''$\,=\,10 to 14)}. The numbers (in black) on the horizontal sections of the curves denote the $v'$ levels of H$_{2}^{+}$.} 
\label{fig:H2_photo_x-sections}
\end{figure}

Figure~\ref{fig:H2_photo_x-sections} shows the state-to-state H$_2^*$($v''$\,$>$\,4) photoionization
 cross sections adopted in our photochemical models for \mbox{reaction~(\ref{reaction_photoi})}.
These cross sections were computed by \cite{Ford75}  and are provided for photoionization, where the product H$_2^+$ is formed in a specific vibrational level $v'$. Hence, the result is a superposition of step functions, each corresponding to a different value of $v'$. These calculations only consider rotational transitions \mbox{$J''$\,=\,1\,$\rightarrow$\,$J'$\,=\,1}.
Here we explicitly calculated the FUV photon energy thresholds ($E_t$)  for each state-to-state process, from \mbox{$v''$\,=\,4 to 14} and leading to vibrationally excited H$_2^+$($v'$). Specifically,
we used the energy of the \mbox{H$_2$($v''$$\geq$4,\,$J''$\,=\,1)} levels \citep{Komasa11} 
and of \mbox{H$_2^+$($v'$,\,$J''$\,=\,1)} levels \citep{Jaquet08,Hilico00}
 and computed:
\begin{equation}
E_t = E_{\rm I}({\rm H_2}) - \Delta E(v'' -v'),
\label{reaction_threshold}
\end{equation}
where $E_{\rm I}$ is the H$_2$ ionization energy \citep[15.426\,eV, e.g.,][]{Shiner93} and
\mbox{$\Delta E(v'' -v')$\,=\,$E_{\rm H_2}$($v''$,\,$J''$=1)\,$-$\,$E_{\rm H_2^+}$($v'$,\,$J'$=1)}.

\begin{figure}[th]
\centering   
\includegraphics[scale=0.5, angle=0]{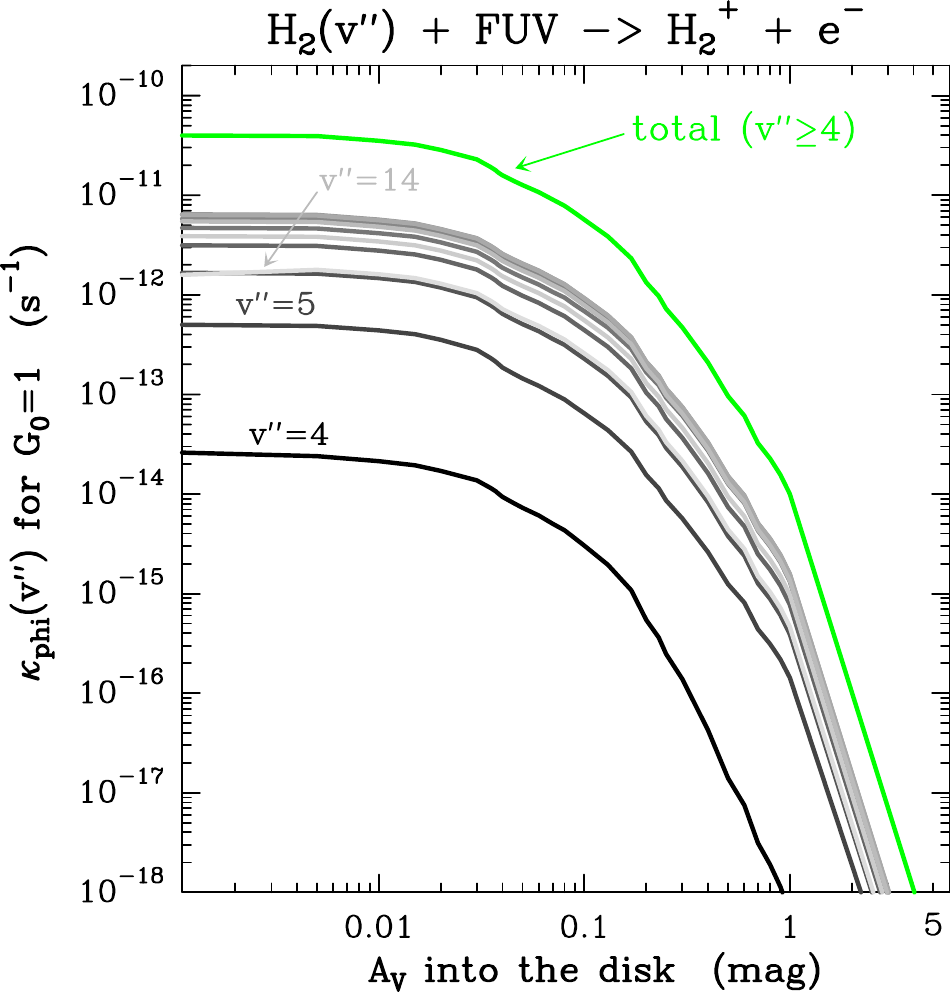}
\caption{Resulting H$_2$($v''$$\geq$4) photoionization rates (normalized to $G_0$\,=\,1) as a function of depth into the disk after direct integration of equation~(\ref{reaction_kappa_phi}) 
using the $\lambda$-dependent FUV field of our model of \mbox{d203-506}.
In this model, the largest contribution to H$_2$($v'' \geq 4$) photoionization comes from the 
$v'' = 10$ level.} 
\label{fig:H2_photo_rates}
\end{figure}

Since we do not explicitly follow
the vibrational excitation of H$_2^+$ molecules, we use
\mbox{$\sigma_{\lambda}$\,($v''$)\,=\,$\sum_{v'}$\,$\sigma_{\lambda}$($v''$$\rightarrow$$v'$)} and integrate these $\sigma_{\lambda}$\,($v''$) sections cross sections over the local
FUV energy density at different disk positions (equation~\ref{reaction_kappa_phi}).
\mbox{Figure~\ref{fig:H2_photo_rates}} shows the resulting rates for the specific
model of \mbox{d203-506} (normalized to $G_0$\,=\,1). To simplify their use in the
Meudon PDR code, we carried out $v''$-state-dependent power-law fits,  of the form
 \mbox{$\kappa_{\rm phi}$($v''$;\,$A_V$)\,=\,$G_0$\,$\kappa_0$($v''$)\,exp\,($-\gamma$($v''$)\,$A_V$)}, to these rates \mbox{(in s$^{-1}$)}.
 The  $\kappa_0$($v''$) and $\gamma$($v''$) parameters
 are tabulated inside the code, and used
 to determine the total H$_2^+$ formation rate due to H$_2^*$($v''$), summed over all $v''$ levels, FUV photoionization.

\clearpage

\section{State-dependent rate constants for reaction
\mbox{H$_{2}$ + HOC$^+$\,$\rightarrow$\,H$_{3}^{+}$ + CO}
and isomerization}
\label{appendix-hocp_h2}

The title reaction has been studied experimentally. \mbox{\cite{Freeman87}} measured a rate constant of $4.7 \times 10^{-10}$\,cm$^3$\,s$^{-1}$ at 300\,K. At a lower temperature of 25\,K, however, \cite{Smith02} did not detect H$_3^+$ formation, observing only the isomerization product HCO$^+$. \cite{Herbst-Woon96} calculated the stationary points on the corresponding potential energy surface, finding a submerged barrier. From this, they proposed that the reaction may proceed via an isomerization route forming \mbox{HCO$^+$\,+\,H$_2$} (a highly exothermic pathway), or alternatively form \mbox{H$_3^+$\,+\,CO}, which is endothermic by 0.14\,eV (including ZPEs). As a result, the latter pathway was considered unimportant.

Most theoretical studies have focused on the HOC$^+$/HCO$^+$ isomerization ratio in the H$_3^+$ + CO reaction \citep{Le-etal:10,Yu:11,Zhu-etal:19,Saito-etal:21}. \cite{Yu:11} investigated the H$_2$ + HOC$^+$ reaction using on-the-fly classical trajectory simulations at a relatively low level of theory (SAC-MP2), and did not report the formation of H$_3^+$ products.

\begin{figure}[th]
\begin{center}
  \includegraphics[width=0.97\linewidth]{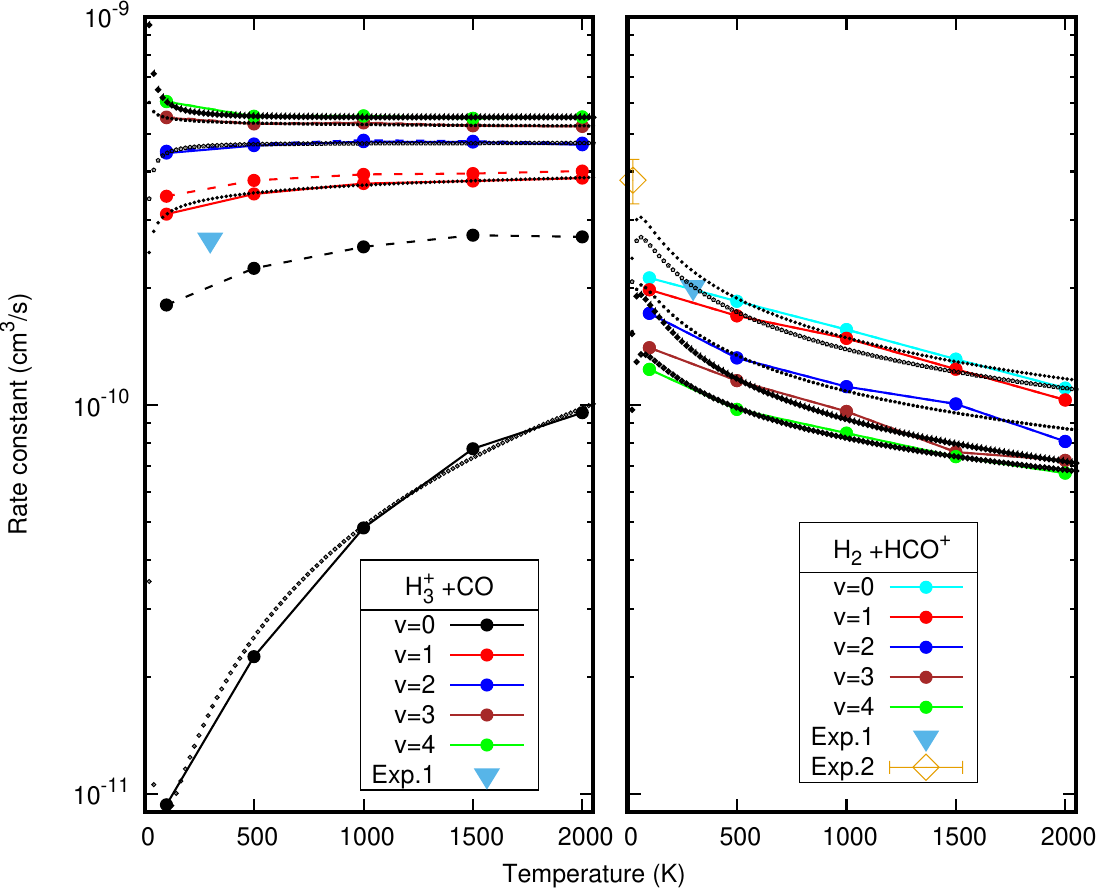}
  \caption{{\mbox{H$_2$($v$ = 0–4, $j$ = 0) + HOC$^+$}, obtained using a QCT method, are shown with and without the ZPE correction for the H$_3^+$ ZPE. Solid lines represent the corrected results; dashed lines show the uncorrected ones. Dotted curves represent the Arrhenius fit to the ZPE-corrected rates. The yellow point in the right panel indicates the experimental rate constant reported by \cite{Smith02}, while the blue triangles represent the rate constants reported by \cite{Freeman87}.}}    
\label{fig:HOC+H2rates}
\end{center}
\end{figure}

In this work, we study the title reaction using the high-level potential energy surface (PES) developed by \cite{Zhu-etal:19}. This PES reveals an endothermic direct pathway from reactants to products, with an energy difference of 0.125\,eV—slightly lower than the value reported by \cite{Herbst-Woon96}. Along the entrance channel, the reaction dynamics can also proceed through a submerged barrier ($-0.085$\,eV including ZPE) toward a deep \mbox{H$_2$--HCO$^+$} well, from which the system evolves to \mbox{H$_2$\,+\,HCO$^+$} products. These products lie approximately 1.75\,eV below the \mbox{H$_2$\,+\,HOC$^+$} reactants.

In order to study the \mbox{H$_2$($v$=0--4,\,$j$=0)\,+\,HOC$^+$\,$\rightarrow$\,H$_3^+$\,+\,CO} reaction, it is important to account for the ZPEs of both reactants and products. Here, we employ a quasi-classical trajectory (QCT) method, as implemented in the MDwQT code \citep{Sanz-Sanz-etal:15}, in which the initial vibrational conditions of the reactants are generated using the adiabatic switching method \citep{Grozdanov-Solovev:82,Qu-Bowman:16}. This approach allows us to include the vibrational energy of H$_2$ in different vibrational levels, as well as that of HOC$^+$ in its ground vibrational state (with a ZPE of 0.336\,eV, consistent with the values reported by \cite{Kraemer-Spirko:10}).

The rate constants obtained in the \mbox{$T$\,$\simeq$\,100\,--\,2{,}000\,K} range are shown in Fig.~\ref{fig:HOC+H2rates} for several initial vibrational states of H$_2$. For each case, two sets of results are presented: one including all reactive trajectories (dashed lines), and another excluding those trajectories that result in product vibrational energies below the corresponding ZPE. The H$_3^+$ formation is strongly affected by this correction. The ZPE-corrected results are reliable because the formation of H$_3^+$ proceeds through a direct mechanism. Thus, selecting trajectories with physically meaningful vibrational energies at both the beginning and the end effectively mitigates potential inaccuracies introduced by classical
mechanics calculations. 

For H$_2(v=0)$, the ZPE-corrected rate constants decrease significantly with decreasing temperature. This trend is consistent with the absence of H$_3^+$ products in the experimental study by \cite{Smith02}. On the other hand, the rate constant measured for isomerization toward HCO$^+$ (yellow point in the right panel of Fig.~\ref{fig:HOC+H2rates}) is in good agreement with the calculated rates.

The value reported by \cite{Freeman87} at 300\,K for H$_3^+$ formation is much higher than our rate constant for initial \mbox{H$_2(v=0)$} but slightly lower than the calculated rates for 
\mbox{H$_2(v=1)$}. To reconcile these differences, we suspect that the HOC$^+$ reactants in the \cite{Freeman87} experiment may have been vibrationally excited. This assumption allows us to provide a coherent interpretation that integrates all available experimental and theoretical data.

\begin{figure}[th]
\begin{center}
  \includegraphics[width=0.661\linewidth]{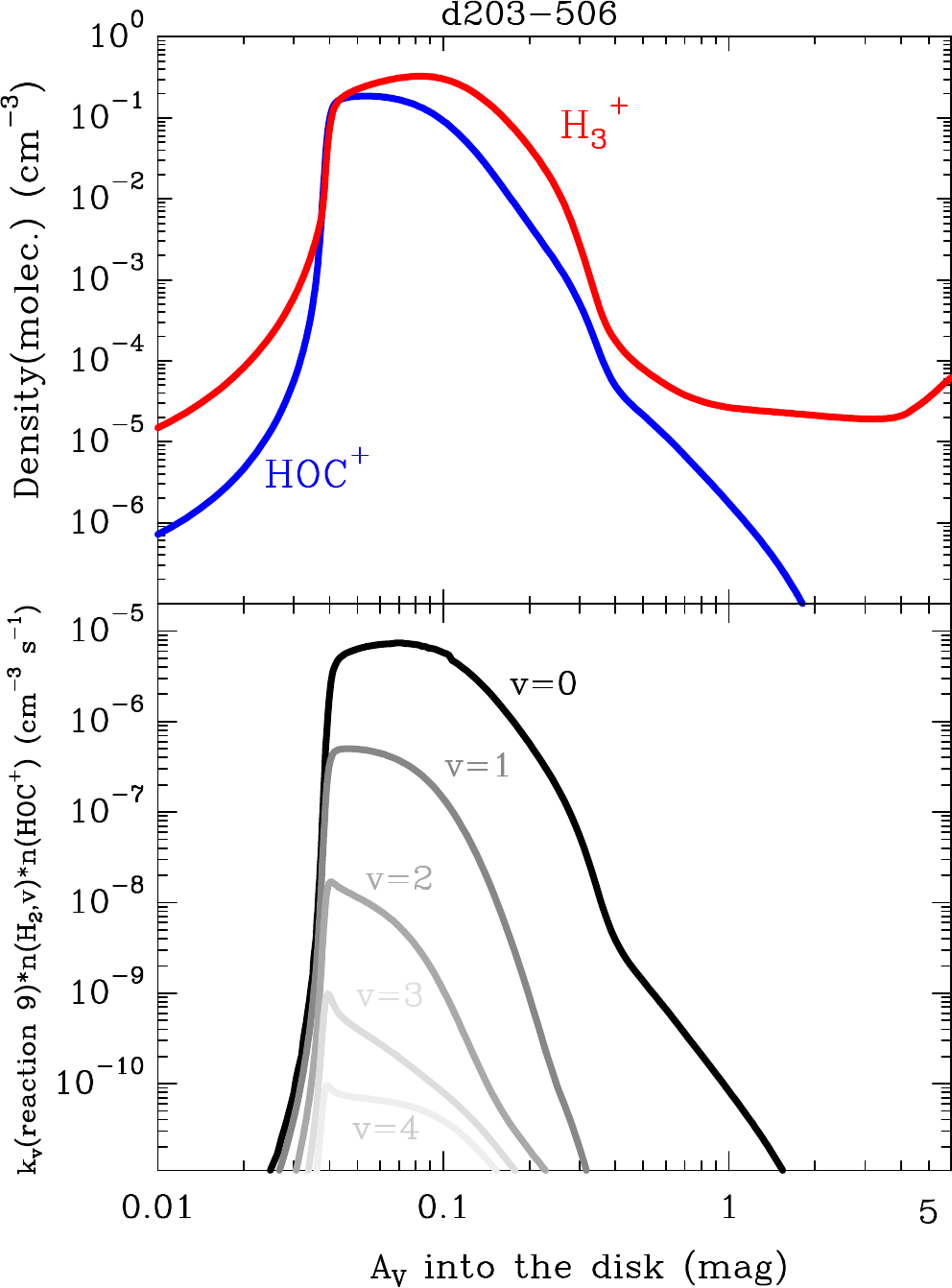}
  \caption{\textit{Upper panel}: H$_3^+$ and HOC$^+$ volume densities  predicted
  by the reference model of \mbox{d203-506}. \textit{Lower panel}: 
  Vibrational-state dependence of the  
  H$_3^+$ formation rate, $F$ in cm$^{-3}$\,s$^{-1}$, for reaction \mbox{H$_2(v)$\,+HOC$^+$\,$\rightarrow$\,H$_3^+$\,+\,CO}, as a function
  of depth into the PDR.}    
\label{fig:HOC+rates_PDR}
\end{center}
\end{figure}

\mbox{Figure~\ref{fig:HOC+rates_PDR}} shows the evolution of the vibrational-state 
dependent H$_3^+$ formation rate, $F$ (in cm$^{-3}$\,s$^{-1}$), for the reaction \mbox{H$_2(v)$ + HCO$^+$ $\rightarrow$ H$_3^+$ + CO}, after incorporating the rate constants from Fig.~\ref{fig:HOC+H2rates} into our reference PDR model.

 \clearpage

\section{Coupled non-LTE excitation and chemical formation model of H$_3^+$}
\label{appendix-grosbeta}

We briefly describe the non-LTE excitation models  carried out with the zero-dimensional (0D) escape-probability radiative transfer model \texttt{GROSBETA} \citep{Black98, Tabone21, Zannese24}, an enhanced version of \texttt{RADEX} \citep{Tak07}. We used this model to test whether the physical conditions and H$_3^+$ chemistry predicted by our  PDR model  can reproduce the IR H$_3^+$ line intensities reported  in \mbox{d203-506} \citep{Ilane25b}. \texttt{GROSBETA} solves the statistical equilibrium equations by accounting for local chemical  formation and destruction rates, ro-vibrational
collisional excitation, spontaneous emission, and line opacities providing physically consistent predictions of the emergent line intensities under non-LTE excitation conditions.

A  proper treatment of H$_3^+$ vibrational excitation must account for the chemical formation and destruction rates within the statistical equilibrium equations that govern the 
ro-vibrational level populations. These equations take the form:
\begin{eqnarray}
\sum_{j>i} n_j\,A_{ji} + \sum_{j \neq i} n_j \left(B_{ji}\,\bar{J}_{\nu} + C_{ji} \right) + F_i =\\
= n_i \left( \sum_{j<i} A_{ij} + \sum_{j \neq i} \left(B_{ij}\,\bar{J}_{\nu} + C_{ij}  \right) \,+\, D_i \right),
\end{eqnarray}
where $n_i$ [cm$^{-3}$] is the population of ro-vibrational level $i$; $A_{ij}$ and $B_{ij}$ are the Einstein coefficients for spontaneous and stimulated emission, respectively; and $C_{ij}$ [s$^{-1}$] is the rate of inelastic collisions. The latter is given by $C_{ij} = \sum_k \gamma_{ij,k}(T)\,n_k$, where $\gamma_{ij,k}(T)$ \mbox{[cm$^3$\,s$^{-1}$]} are the temperature-dependent collisional rate coefficients, and $k$ denotes collision partners H$_2$, H, or $e^-$. The term $\bar{J}_{\nu}$ represents the mean intensity of the radiation field averaged over the line profile and solid angle.
In the above expression, $F_i$ is the formation rate into level $i$ per unit volume,  and $n_i\,D_i$ is the destruction rate from level $i$ per unit volume, both
in \mbox{cm$^{-3}$\,s$^{-1}$}. Assuming a level-independent destruction rate ($D_i = D$), one can model the ro-vibrational level distribution of newly formed H$_3^+$ molecules as a Boltzmann distribution characterized by a formation temperature $T_{\rm form}$, where
\begin{equation}
F_i = F\, g_i \, \frac{e^{-E_i / kT_{\rm form}}}{Q(T_{\rm form})},
\end{equation}
and $F$ [cm$^{-3}$\,s$^{-1}$] is the total (state-averaged) formation rate per unit volume, $g_i$ is the degeneracy of level $i$, and $Q(T_{\rm form})$ is the H$_3^+$ partition function at the formation temperature $T_{\rm form}$.

As  these are constant column density $N$(H$_3^+$) excitation models, we used a normalized formation rate \mbox{$F$\,=\,$\sum F_i$} that adopts \mbox{steady-state} H$_3^+$ abundances. 
That is, \mbox{$F$\,=\,$\sum F_i$\,=\,$x$(H$_3^+$)\,$n_{\rm H}$\,$D$} \mbox{$[\rm cm^{-3} s^{-1}]$}, where $x$ refers to the H$_3^+$ abundance with respect to H nuclei, and $D$ is the chemical
destruction rate in s$^{-1}$.

\subsection{Model input}

We included 748 H$_3^+$ vibrational-rotational levels with $J$\,$<$\,9, except for the (9,9) metastable level, and nearly 
29{,}000
radiative transitions. The spectroscopic data have been taken from the ExoMol MiZATeP data set
\citep{Mizus17,Bowesman23}.  The \mbox{H$_2$\,+\,H$_3^+$} and \mbox{H\,+\,H$_3^+$} inelastic rotational collision  rates are taken from \citet{Gomez-Carrasco12} and \citet{Felix-Gonzalez25}, respectively.
The inelastic electron collision rates are taken from \citet{Kokoouline10} plus our own Coulomb-Born rate approximation for the vibrational transitions. In addition, we have included some approximate rate coefficients for \mbox{H--H$_3^+$} and 
\mbox{H$_2$--H$_3^+$}  vibrational collisional excitation. 
The typical magnitude of the relevant de-excitation rate coefficients is  \mbox{$\gamma_{ij}$\,$\gtrsim 10^{-11}$ cm$^3$\,s$^{-1}$}.

The input parameters are representative of \mbox{zone~$ii$} in the disk PDR: 
\mbox{$n$(H$_2$) = $n$(H) = 5$\times$10$^6$\,cm$^{-3}$}, \mbox{$n$(e)\,=\,300\,\,cm$^{-3}$},
\mbox{$T_{\rm gas}$ = 1{,}000\,K}, and 
\mbox{$N$(H$_3^+$) = 10$^{13}$\,cm$^{-2}$}. 
Based on our PDR model results, H$_3^+$ formation is assumed to be dominated by reaction~(\ref{reaction_hocp}) 
when H$_2$ is in its ground vibrational state ($v''$\,=\,0). 
Therefore, we adopt a  formation rate of \mbox{$F$(H$_3^+$)\,$\simeq$\,$k_9$($v''$=0)\,$\cdot$\,$n$(H$_2$;\,$v''$=0)\,$\cdot$\,$n$(HOC$^+$)
$\simeq$\,10$^{-5}$\,cm$^{-3}$\,s$^{-1}$} (see detailed predictions in Fig~\ref{fig:HOC+rates_PDR}).
Thus, the  chemical destruction rate of H$_3^+$ is
 \mbox{$D$\,= $F$ / $x$(H$_3^+$)\,$\cdot$\,$n_{\rm H}$\,$\simeq$\,10$^{-4}$\,s$^{-1}$} (or a lifetime of 2.8\,h),
 adopting \mbox{$x$(H$_3^+$)\,=\,10$^{-8}$} and \mbox{$n_{\rm H}$\,=\,10$^7$\,cm$^{-3}$}.

Finally,  we adopted an intrinsic line widths of 2.7\,km\,s$^{-1}$ as indicated by velocity-resolved ALMA observations of \mbox{d203-506} \citep{Berne24,Goico24}, and obtained the synthetic line flux by adopting a source emitting area with a solid angle equivalent to an aperture of 0.1$''$, consistent with the line fluxes extracted by \citet{Ilane25b} from JWST/NIRSpec observations.

\subsection{Model results}

Adopting a formation temperature of 
$T_{\rm form}$\,=\,3{,}000\,K\,$\approx$ $E_{\rm H_2}(v'', J'')/k - 1{,}500$\,K--corresponding to H$_3^+$ being primarily formed via reactions between \mbox{H$_2$($v''$=0;\,$J''$$\geq$7)} and 
HOC$^+$--results in a non-LTE excitation model that is consistent the intensities of the unblended H$_3^+$ 
\mbox{ro-vibrational} lines reported by  \citet{Ilane25b} to within a factor of about two. 
Table~\ref{Table_intensities} shows the output of this excitation model.

In addition, we also find implications for the H$_3^+$ forbidden pure rotational emission spectrum, which has yet to be observed in the ISM.
Due to its symmetric, equilateral triangular structure, H$_3^+$ lacks a permanent electric dipole moment. Ordinarily, this would imply that pure rotational transitions are strictly forbidden. However, the symmetry of H$_3^+$ is slightly broken by rotation-vibration coupling, which induces a weak dipole moment. As a result, radiative transitions between rotational levels that comply with selection rules are permitted, though they are very slow. 
Our model suggests that the most intense transition of H$_3^+$ in the parameter space of interest is the pure rotational transition \mbox{$(J,K)=(5,0)-(4,3)$} at  16.325\,$\mu$m. 
The transition frequency has been accurately predicted from measured combination differences by
\citet{Markus19}.  
Our non-LTE model predicts an intensity that is rather close to the 3$\sigma$ detection  limit  on integrated intensity in the JWST-MIRI spectrum of \mbox{d203-506}. In other words, the column density of H$_3^+$ at 
$T_{\rm gas}$\,$\simeq$1000\,K cannot be much larger than 10$^{13}$\,cm$^{-2}$, otherwise this 16.325\,$\mu$m line should be detected above the strong continuum.

\end{appendix}

\end{document}